\documentclass[twocolumn,journal]{IEEEtran}
\usepackage{amssymb}
\usepackage{amsmath,amsfonts,amssymb}
\usepackage{cite}
\usepackage{epsfig}
\usepackage{graphicx}
\usepackage{subfigure}
\usepackage{color}

\usepackage{url}
\usepackage{longtable,rotating,multirow,multicol}
\usepackage{setspace}
\usepackage{bm}
\usepackage{threeparttable}
\usepackage{lineno}
\renewcommand{\baselinestretch}{1.0}

\newtheorem{theorem}{\textbf{Theorem}}
\newtheorem{corollary}{\textbf{Corollary}}

\newtheorem{lemma}{\textbf{Lemma}}

\newtheorem{remark}{\textbf{Remark}}
\newtheorem{problem}{\textbf{Problem}}
\begin{document}
\title{Detection and Isolation of Wheelset Intermittent Over-creeps for Electric Multiple Units Based on a Weighted Moving Average Technique}
\author{
	\vskip 1em
	Yinghong~Zhao, \emph{Student Member,~IEEE},
    Xiao~He, \emph{Member,~IEEE},
    Donghua~Zhou, \emph{Fellow,~IEEE},
	and Michael~G.~Pecht, \emph{Fellow,~IEEE}

	\thanks{
		This work was supported by the National Natural Science Foundation of China (NSFC) under Grants 61751307, 61733009, the Research Fund for the Taishan Scholar Project of Shandong Province of China (LZB2015-162), and the Key Project from Natural Sciences Foundation of Guangdong Province under Grant 2018B030311054. (Corresponding author: Donghua Zhou.)

    Yinghong Zhao and Xiao He are with the Department of Automation, BNRist, Tsinghua University, Beijing 100084, China. Yinghong Zhao is also with the Center for Advanced Life Cycle Engineering (CALCE), University of Maryland, College Park, MD 20742, USA. (e-mail: zyh14@mails.tsinghua.edu.cn; hexiao@tsinghua.edu.cn).

	Donghua Zhou is with the College of Electrical Engineering and Automation, Shandong University of Science and Technology, Qingdao 266590, China, and also with the Department of Automation, BNRist, Tsinghua University, Beijing 100084, China. (e-mail: zdh@mail.tsinghua.edu.cn).

    Michael G. Pecht is with the Center for Advanced Life Cycle Engineering (CALCE), University of Maryland, College Park, MD 20742, USA. (e-mail: pecht@umd.edu).
	}
}
\maketitle
\begin{abstract}
Wheelset intermittent over-creeps (WIOs), i.e., slips or slides, can decrease the overall traction and braking performance of Electric Multiple Units (EMUs). However, they are difficult to detect and isolate due to their small magnitude and short duration.
This paper presents a new index called variable-to-minimum difference (VMD) and a new technique called weighted moving average (WMA). Their combination, i.e., the WMA-VMD index, is used to detect and isolate WIOs in real time. Different from the existing moving average (MA) technique that puts an equal weight on samples within a time window, WMA uses correlation information to find an optimal weight vector (OWV), so as to better improve the index's robustness and sensitivity. The uniqueness of the OWV for the WMA-VMD index is proven, and the properties of the OWV are revealed. The OWV possesses a symmetrical structure, and the equally weighted scheme is optimal when data are independent. This explains the rationale of existing MA-based methods. WIO detectability and isolability conditions of the WMA-VMD index are provided, leading to an analysis of the properties of two nonlinear, discontinuous operators, $\min$ and $\textrm{VMD}_i$. Experimental studies are conducted based on practical running data and a hardware-in-the-loop platform of an EMU to show that the developed methods are effective.
\end{abstract}
%Wheel and rail are the fundamental elements of every rail vehicle. The interaction of both constitutes the wheel/rail system, which has multiple functions.

\section{Introduction}\label{IntroductionSec}
Electric multiple units (EMUs) such as high-speed trains (HSTs) \cite{Chen2018Deep,Zhou2018Fault,Chen2019Broad}, electric commuter trains (ECTs) \cite{Kadowaki2007Antislip,Ohishi2009Realization}, and modern urban rail transits (URTs) have become an important and indispensable part of public transport systems \cite{Zheng2013China}. Being able to accomplish long-distance and high-speed transportation makes HSTs much easier to connect different cities \cite{Chen2019Review}. ECTs and URTs primarily operate within a city to transport large numbers of people at higher frequency over short distances. The EMU is popular around the world due to its superior traction and braking (TB) performance.

The TB performance of EMUs strongly depends on an adhesion force arising between wheelsets and rails (WRs).
Modern studies \cite{Wickens2003Fundamentals,Iwnicki2006Handbook,Knothe2017Rail} on creep theory of WR systems have shown that, when the load and the WR surface condition are constant, creep velocity (relative velocity between the WR) is the main factor that affects the adhesion force. The creep velocity should be controlled within a certain range to track a high adhesion point \cite{Diao2017Taking}.
While appropriate creep is desired, an over-creep (i.e., slip or slide) has several disadvantages, such as a decrease in TB performance and additional wear to WRs.
Large over-creeps can sharply decrease the adhesion force, severely damage the WRs, significantly reduce the system securities, and even cause wheelset derailments.
Moreover, the resulted vibrations, noises and re-adhesion processes discomfort the passengers.
Therefore, monitoring the creep phenomenon, and detecting and isolating the over-creep in real time are necessary.
%As the creep velocity continues increasing and exceeds a certain value, a slip or slide takes place.

Over-creeps occur for many reasons such as low WR adhesion conditions caused by high humidity, dust, rain, frost, snow or ice. Contaminants between contact surfaces (e.g., decomposing leaves or oily substances), as well as a harsh working environment (vibration, shock or electromagnetic interference) can lead to over-creeps. Moreover, it is difficult to find controller parameters that meet all WR conditions \cite{Diao2017Taking}. As a result, unsuitable controller parameters may bring about over-creeps when faced with varied working conditions. Other reasons include the wear or damage of WRs (e.g., due to fatigue or over-creeps), rail irregularities, and component degradation of railway systems (sleepers or rail fastenings) or TB systems (transformers, inverters, motors or brake cylinders) \cite{Knothe2017Rail}.

A main difficulty in the detection and isolation (DI) of the over-creep is that it occurs intermittently. It has the characteristics of intermittent faults, e.g., lasting a limited period of time and then disappearing \cite{Zhou2019Review}.
Since over-creeps are serious threats to EMU security, anti-slip/slide (ASS) systems have always been the core part of EMUs. In practice, each axle of the EMU is equipped with a separate speed sensor. Once the velocity difference between wheelsets or wheelset acceleration/deceleration (WAD) goes beyond its preset values, the corresponding wheelset is considered to be slipping/sliding \cite{Zhang2017Rail}.
Although these ASS strategies can ensure the safe operation of EMUs by detecting and isolating large over-creeps timely, they are not effective for some intermittent over-creeps (IOs). These IOs are not large enough to trigger an alarm in current ASS systems, but they still impair the TB performance. To overcome this problem, some efforts have been made to improve the current ASS strategies.
%Currently, strategies of creep monitoring adopted by ASS systems of EMUs are relatively simple in practice.

Since slips decrease the load torque of traction motors (because of smaller adhesion coefficients), a multi-rate extended Kalman filter (MREKF) was constructed using an induction motor model to estimate the load torque in real time \cite{Wang2016Locomotive}. Slips were then detected from the change of estimated load torque. The MREKF only uses voltage and current information of traction motors, and thus is a speed sensorless slip detection method.
An alternative method to estimate the load torque and adhesion coefficient was proposed by \cite{Kadowaki2007Antislip,Ohishi2009Realization}, where estimations were given by a disturbance observer. The disturbance observer was constructed based on the TB model of railway vehicles (the motion equation of EMUs, torque equation of electric motors, rotation equation of wheelsets, etc.), and a secondary flux-based angular speed estimator. Instead of the load torque, they used the differential value of the estimated wheelset velocity to detect over-creeps \cite{Kadowaki2007Antislip,Ohishi2009Realization}. The method has shown a desired performance in its application to an ECT (Series 205-5000 of the East Japan Railway Company) \cite{Kadowaki2007Antislip}. Subsequently in \cite{Ohishi2009Realization}, to reduce the adverse effects of bogie vibrations, a high-order disturbance observer considering the resonant frequency of bogie systems was proposed.
Moreover, the use of torsional vibration for slip detection was reported in \cite{Mei2009A}. Torsional vibration of wheelsets was estimated by a Kalman filter (KF) via wheelset dynamics.
Since a prior knowledge of various models is required, these methods can be referred to as model-based methods.
%These methods use a disturbance observer or kalman filter to obtain the current motor load torque according to measurable variables such as voltage and current.
%When the wheelset speed is available, this idea has also been adopted by. enhance the performance. In one of these studies. Improve the results of pervious research. In place of the speed information. But this approach by our opinion, does not allow reaching the realization of maximum possible adhesion. By focusing on the current of each motor.

As for data-driven methods, several researchers \cite{Kim2015Slip,Spiryagin2008Control,Tu2017A,Cai2009A,Watanabe2002Basic,Yamashita2005Readhesion} detected over-creeps by integrating different sensor information.
In \cite{Kim2015Slip}, multiple sensors, such as tachometer, differential GPS, inertial navigation and RFID, were fused by a two-stage federated KF (TS-FKF) to estimate the velocity of the EMU and then detect over-creeps.
In \cite{Spiryagin2008Control}, directional microphones were utilized to scan the rolling noises of wheelsets, and the spectrum of the rolling noise was used to detect over-creeps.
In \cite{Tu2017A}, measurements from the odometer, Doppler radar and accelerometer were utilized by an adaptive fuzzy algorithm to detect over-creeps.
In \cite{Cai2009A}, velocity estimations from the GPS receiver and odometer were compared to detect over-creeps.
In \cite{Watanabe2002Basic,Yamashita2005Readhesion}, differences between motor currents were employed to detect over-creeps. Note that for these methods, additional sensors are needed for ASS systems.
Among over-creep DI methods using only speed information \cite{Saab2002Compensation,Cimen2018A,Xiao2003Locomotive,Huang2007Optimized,Yasuoka1997Improvement,Niu2019Fault}, \cite{Saab2002Compensation} examined the variation of vehicle accelerations, which can be estimated by a steady-state KF using tachometer data. In \cite{Cimen2018A}, the WAD obtained by differentiating the measured wheelset velocity was utilized to detect over-creeps. In \cite{Xiao2003Locomotive,Huang2007Optimized}, both the WAD and the discrete differential value of WAD were used to detect over-creeps after they were denoised by the wavelet transform. In \cite{Yasuoka1997Improvement}, the WAD and the velocity difference between the wheelset and vehicle were utilized to detect over-creeps. In traction mode, the vehicle velocity was estimated on the basis of the lowest wheelset velocity and the vehicle's inertia.
Most recently, in \cite{Niu2019Fault}, slides were detected and diagnosed among several faults with the help of a Petri net model, wherein only the wheelset velocities were used.

With the development of modern control (e.g., adhesion control \cite{Diao2017Taking} and fault-tolerant control \cite{Mao2019Adaptive}) technologies of EMUs, IOs that have small magnitude and short duration are more likely to occur. Several above-mentioned model-based methods or multi-sensor-based data-driven methods have been reported to improve the over-creep DI performance. Data-driven methods without utilizing additional sensors still need further study. Therefore, the objective of this paper is to improve the DI performance of wheelset IOs (WIOs) using only the mounted speed sensors. Main contributions of this paper can be summarized as follows:
1) Inspired by the current ASS strategies, a new index called variable-to-minimum difference (VMD) for the DI of anomalies is developed. It is useful in many real-world applications.
2) To improve its robustness and sensitivity, a weighted moving average (WMA) technique is presented. Different from the existing MA technique that puts an equal weight on samples within a time window, WMA uses correlation information to find an optimal weight vector (OWV). The uniqueness of the OWV for the WMA-VMD index is proven.
3) Methods to determine the OWV are given, and properties of the OWV are discussed. We reveal that the OWV possesses a symmetrical structure, and an equally weighted scheme is optimal when data are independent. This explains the rationale of existing MA-based methods.
4) WIO detectability and isolability analyses of the WMA-VMD index are provided, leading to the property analyses of two nonlinear, discontinuous operators, $\min$ and $\textrm{VMD}_i$.
5) Experimental studies are carried out, which demonstrate the effectiveness of the proposed methods. Experimental results are discussed, with comparison to current ASS strategies on EMUs.
%The ASS systems of EMUs are requested to have advanced over-creep DI systems.

Section \ref{BackgroundSec} introduces the creep and over-creep phenomena, currently used ASS strategies, and the objective of this paper.
Then, the WMA-VMD index is proposed in Section \ref{MethodologySec} to address the WIO detection and isolation issue.
Detectability and isolability analyses are provided in Section \ref{DIanalysisSec}. Experimental studies are presented in Section \ref{ExperimentSec}. The paper is concluded in Section \ref{ConclusionSec}.

{\bf Notation:} Except where otherwise stated, the notations used throughout the paper are standard.
${\mathbb E}\{x\}$ and ${\mathbb V\rm{ar}}\{x\}$ stand for the expectation and variance of a random variable $x$, respectively;
${\mathbb C\rm{ov}}\{x,y\}$ represents the covariance between random variables $x$ and $y$.
${\mathbb R}^{n}$ and ${\mathbb R}^{n \times m}$ denote the $n$-dimensional Euclidean space and the set of all $n \times m$ real matrices.
$\mathbf{A}^T$, $\mathbf{A}^{-1}$, $|\mathbf{A}|$ and \textrm{adj}($\mathbf{A}$) stand for the transpose, the inverse, the determinant and the adjoint of a matrix $\mathbf{A}$, respectively.
$\nabla_{\mathbf{a}_W}{\cal L}(\mathbf{a}_W,\lambda)$ is the gradient of ${\cal L}$ with respect to $\mathbf{a}_W$.
$\nabla^2_{\mathbf{a}_W}{\cal L}(\mathbf{a}_W,\lambda)$ is the Hessian matrix of ${\cal L}$ with respect to $\mathbf{a}_W$.
Scalars $a_1\cdots a_W$ form a row vector by $[a_1,a_2,\cdots,a_W]$, and form a column vector by $[a_1;a_2;\cdots;a_W]$. $\triangleq$ is to give definition.
$H_{l,l'}$ or $[\mathbf{H}]_{l,l'}$ is an element of matrix $\mathbf{H}$ located in the $l$th row and $l'$th column.
$\mathbf{A}_{\backslash i\backslash j}$ is the matrix obtained from $\mathbf{A}$ by deleting the row and column containing $A_{i,j}$.
$\mathbf{I}_{p}$ and $\mathbf{e}_{pi}$ denote the $p$-dimensional identity matrix and its $i$th column, respectively; $\mathbf{1}_W$ and $\mathbf{0}_W$ denote the $W$-dimensional column vectors with all of its entries being one and zero, respectively.
$\mathbf{A}\prec\mathbf{B}$ and $\mathbf{A}\preceq\mathbf{B}$ mean that $\mathbf{A}-\mathbf{B}$ is negative definite and negative semidefinite, respectively.

\section{Creep, current strategy, and objective}\label{BackgroundSec}
In this section, the creep and over-creep phenomena, as well as the currently used ASS strategies on EMUs are briefly described. Then, the objective of this paper is presented.

\subsection{Creep and over-creep phenomena}\label{CreepSub}
\begin{figure}
\centering\includegraphics[width=1\linewidth]{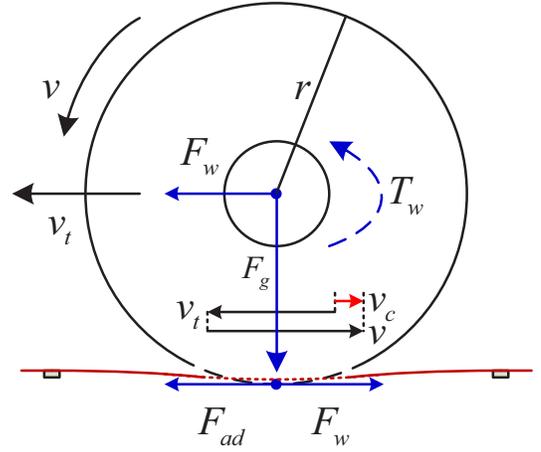}
\caption{An illustration of the creep phenomenon.} \label{CreepPhenomenon}
\end{figure}

The creep phenomenon can be defined as a wheel-rail micro-elastic slide through relative deformation \cite{Diao2017Taking}. As shown in Fig.~\ref{CreepPhenomenon}, since both the wheel and rail are elastic bodies, a normal load $F_g$ leads to local micro-elastic deformation at their contact point. Then, an area of contact (i.e., the contact patch) is formed.
As a result, when a tractive torque $T_w$ or braking torque is applied, pure rolling rarely takes place and a slight slide arises between the wheel and rail, called creep.
The creep causes a small difference between the forward velocity $v_t$ and wheel circumference line velocity $v$. This relative difference is called creep velocity $v_c$ and is given by
\begin{align}\label{CreepVelocity}
v_{c}=v-v_t.
\end{align}
The adhesion coefficient $\eta$, which is defined as the ratio between adhesion force $F_{ad}$ and normal load $F_g$ as follows,
\begin{align}\label{AdhesionCoefficient}
\eta(v_c)=\frac{F_{ad}(v_c)}{F_g},
\end{align}
has a close relation to $v_c$ when the WR surface condition is constant. Generally speaking, a dry, clean WR surface generates a higher $\eta$ compared with a wet surface. $\eta$ can be further reduced on an oily surface. The $\eta-v_c$ relationships, i.e., adhesion characteristic curves (ACCs), with various WR surface conditions were shown in \cite{Wang2016Locomotive,Diao2017Taking}. ACCs rise first and then decline.
Hence, each ACC has its own peak point $v^*_c$, where the adhesion force $F_{ad}$ reaches a maximum $F_{ad}(v^*_c)=\eta(v^*_c)F_g$.
It can be seen that a proper creep velocity is beneficial to generate the adhesion force, and thus, the creep velocity should be controlled within a certain range to track a high adhesion point \cite{Diao2017Taking}. While appropriate creep is desired, over-creep has several disadvantages, such as reduction in adhesion force and additional wear to WRs.
%\begin{figure}
%\centering\includegraphics[width=1\linewidth]{AdhesionCharacteristicCurves}
%\caption{Adhesion characteristic curves with various WR surface conditions.} \label{AdhesionCharacteristicCurves}
%\end{figure}
%The $\eta-v_c$ relationships, i.e., adhesion characteristic curves (ACCs) with various WR surface conditions are shown in Fig.~\ref{AdhesionCharacteristicCurves}.

There are two kinds of over-creep phenomena, namely, slip and slide.
Since tractive torque $T_w$ is equivalent to the moment of a couple $F_w$, we now use $F_w$ instead of $T_w$ in Fig.~\ref{CreepPhenomenon}.
It can be seen that $F_w$ acting on the contact patch is resisted by a force of friction called adhesion force $F_{ad}$.
When $F_w$ is less than the maximum available adhesion force $F_{ad}(v^*_c)$, their resultant is zero and the remained $F_w$ acting on the wheelset mass center accelerates the EMU.
However, when $F_w$ exceeds $F_{ad}(v^*_c)$, their resultant cannot be zero, resulting in a toque $T_c=(F_w-F_{ad})r$ that rotates the wheelset. In this way, slip takes place and the wheelset accelerates abnormally. Similarly, when the equivalent couple generated by a braking torque exceeds the maximum available adhesion force, slide takes place and the wheelset decelerates abnormally. The result is that the velocity of the slipping (sliding) wheelset becomes greater (lower) than that of other normal wheelsets.

The fact that ACCs rise first and then decline is due to structural changes of the contact patch.
When the WRs have smooth surfaces and constant curvature in the vicinity of the contact patch, the contact patch is believed to be elliptical in shape, and divided into adhesion area and slip/slide area \cite{Wickens2003Fundamentals}. The adhesion area, in which the surfaces are locked together, locates at the front of the contact patch. It reduces progressively as the adhesion force increases. When only the slip/slide area is left, the adhesion between wheel and rail breaks down, leading to the decline in the ACC.

\subsection{Current anti-slip and anti-slide strategies on EMUs}\label{CurrentStrategySub}
Typically, one EMU car has four wheelsets, each equipped with a speed sensor. Thus, a total of four channels of wheelset velocity $\{v_{i}, i=1 \thicksim 4\}$ are measured and utilized by the ASS system. Denote the velocity and acceleration of the $i$th wheelset as $v_{i}$ and $a_{i}=\dot{v}_{i}$, respectively.
Note that $v_t$ is not directly measurable, so currently used ASS strategies are generally based on the wheelset velocity difference and wheelset acceleration as follows.

\subsubsection{In traction mode} the $i$th wheelset is considered to be slipping if $e^p_{i}>J^p_{e}$ or $a_{i}>J^p_{a}$, where
\begin{align}\label{TVDC}
e^p_{i}=v_{i}-\min(v_{1},v_{2},v_{3},v_{4}),
\end{align}
and $J^p_{e},J^p_{a}$ are two preset values.
\subsubsection{In braking mode} the $i$th wheelset is considered to be sliding if $e^b_{i}>J^b_{e}$ or $a_{i}<-J^b_{a}$, where
\begin{align}\label{BVDC}
e^b_{i}=\max(v_{1},v_{2},v_{3},v_{4})-v_{i},
\end{align}
and $J^b_{e},J^b_{a}$ are two preset values.

The if-then logic $e^p_{i}>J^p_{e}$ and $e^b_{i}>J^b_{e}$ are referred as \emph{velocity difference criteria} in traction and braking mode, respectively. Likewise, the if-then logic $a_{i}>J^p_{a}$ and $a_{i}<-J^b_{a}$ are referred as \emph{acceleration criteria} in traction and braking mode, respectively.
Note that the DI of over-creeps is accomplished here in one step. This inspires us to develop the VMD index in Section \ref{MethodologySec}, which can be used in many other practical problems as well.
%which can be further applied to many other practical problems.

\subsection{Objective}\label{ObjectiveSub}
Velocity information plays an important role in the DI of over-creeps. In practice, speed sensors are equipped on bogies, which are in direct contact with rails. As a result, speed sensors are susceptible to noises and disturbances caused by the harsh working environment (e.g., vibration, shock, electromagnetic interference, severe weather). Therefore, modern EMUs must have robust and advanced ASS systems. Although current strategies fully consider the creep mechanisms and consequently can ensure the safe operation of EMUs, their efficiency for WIOs that have small magnitude and short duration remains to be improved.
Therefore, the objective of this paper is to improve the DI performance of WIOs using only the velocity measurements $v_{i}$.
The DI of a WIO means that slip and slide should be detected and distinguished from each other. Moreover, its location, namely, which wheelsets it occurs on, should be determined.

%On the one hand, historical data under normal operating conditions are not fully utilized. Note that the thresholds $J^p_{e},J^p_{a}$ are preset, which means that they should apply to a wide range of WR, weather and working conditions. Thus, the worst condition (e.g., a very oily condition) has to be considered sometimes to ensure the safety \cite{Xu2016Analysis}. This, however, sacrifices the optimal performance. On the other hand, the real-time wheelset velocities are under-utilized. Note that for the VDC, only the latest sample (latest velocities of four wheelsets) is utilized each time to determine whether the wheelsets have over-creeps or not. Inference with merely one sample has a strong uncertainty because of the disturbances. Moreover, considering the harsh working conditions, slight over-creeps can be easily masked by noises and fluctuations. In this way, the detection and isolation indices are hardly sensitive to slight over-creeps and robust to disturbances.

%ʵÑéÊý¾Ý¿ì£¬·ÂÕæÊý¾ÝÄܼì²â£¬µÍËÙÇé¿ö±§ËÀ
%So we aim to improve the sensitivity of velocity difference criterion.
%The sensitivity of velocity difference criterion is much less than the acceleration criterion. Using wheel acceleration of the axles gives superior results.
%rather than a latest series of wheelset velocities. rather than incorporating wheelset velocities from multiple consecutive time instances.
\section{Detection and isolation methodology}\label{MethodologySec}
In this section, inspired by currently used ASS strategies, a VMD index for the DI of anomalies is developed. It can be extended to many real-world applications. Moreover, to improve its robustness and sensitivity, a WMA-VMD index is developed. Methods to determine the OWV are given, and properties of the OWV are discussed to help us better understand the developed index. Finally, the use of WMA-VMD index to detect and isolate WIOs is summarized in an algorithm.
% maximum-based excess. Since, the results can easily be reformulated for sample minima.
% It can be used in many real-world applications.

\subsection{Variable-to-minimum difference index}\label{VMDindexSub}
Without loss of generality, we assume that there are $p$ channels of measurement available $\{v_{i}, i=1 \thicksim p\}$.
For a sample vector $\mathbf{v}=[v_1, v_2,\cdots,v_p]^T\in{\mathbb R}^{p}$, its variable-to-minimum difference (VMD) index for the $i$th variable is defined as
\begin{align}\label{VMD}
\textrm{VMD}_i(\mathbf{v})=v_{i}-\min(v_{1},v_{2},\cdots,v_p).
\end{align}
According to \eqref{TVDC}, the $\textrm{VMD}_i(\mathbf{v})$ index can be used to detect and isolate the $i$th wheelset's slip in traction mode. Moreover, following
\begin{align}\label{MinConvertMax}
\min(-v_{1},-v_{2},\cdots,-v_p)=-\max(v_{1},v_{2},\cdots,v_p),
\end{align}
we have
\begin{align}\label{}
\textrm{VMD}_i(-\mathbf{v})&=-v_{i}-\min(-v_{1},-v_{2},\cdots,-v_{p})\nonumber\\
&=\max(v_{1},v_{2},\cdots,v_{p})-v_{i}.
\end{align}
Then, according to \eqref{BVDC}, the $\textrm{VMD}_i(\mathbf{-v})$ index can be used to detect and isolate the $i$th wheelset's slide in braking mode.

\begin{figure}
\centering\includegraphics[width=1\linewidth]{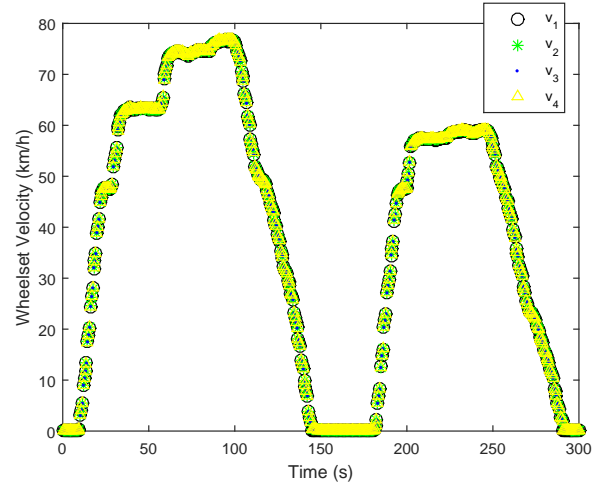}
\caption{Illustration of the wheelset velocities under normal conditions.} \label{PartTrainData}
\end{figure}

The mechanism is readily comprehensible. It can be seen from equation \eqref{CreepVelocity} that a wheelset's absolute velocity equals the train velocity plus its creep velocity. Since all the wheelsets on the same EMU car are controlled by one traction control unit (TCU) and are of the same type and size, they provide almost equal adhesion force under normal conditions.
Considering the EMU's mass is distributed nearly equally on each wheelset whose WR surface conditions are the same, according to \eqref{AdhesionCoefficient}, the creep velocities and consequently the absolute velocities of these wheelsets should be quite similar.
To illustrate this point intuitively, the velocity values of four wheelsets are collected from a practical EMU car under normal conditions and are shown in Fig.~\ref{PartTrainData}. For clarity, only a part of the samples are displayed. This EMU, or specifically, this URT, has made two stops in five minutes. A zero wheelset velocity means that the train has arrived at a station. The inter-station run of URTs usually contains three modes, namely, traction, coasting and braking. It can be observed that the wheelset velocities are almost the same throughout the three modes.

For a sample vector $\mathbf{v}$, the $\textrm{VMD}_i$ index measures the degree of difference between its $i$th variable and its minimum variable. Since these variables should be consistent with each other under normal conditions, all $p$ VMD indices of a normal sample vector should be small. A large $\textrm{VMD}_i(\mathbf{v})$ indicates potential IOs of the $i$th wheelset. This, combined with the diagnosis logic that slip (slide) can only happen in traction (braking) mode, helps us discriminate slip from slide.

The VMD index is applicable to cases where the measurement variables are similar to each other under normal conditions. Note that this kind of situation holds for many practical systems such as brake cylinder systems \cite{Zhou2018Fault}, air brake systems \cite{Guo2019Fault} and bearing systems \cite{Fang2018Multi} of EMUs. Moreover, when this application condition is satisfied, the VMD index has several advantages. First, by using the VMD index, a non-stationary process is transformed into a stationary process, which is easier to be monitored. Second, the DI task can be accomplished by the VMD index in one step because of its clear and readily comprehensible diagnosis logic.
Third, $\textrm{VMD}_i(\mathbf{v})$ is unidimensional and thus easy to be implemented online. Note that it should not be considered as an univariate monitoring strategy because $\textrm{VMD}_i(\mathbf{v})$ makes use of multidimensional measurement.
To utilize the VMD index, we collect $N$ training samples $\{\mathbf{v}_k, k=1,2,\cdots,N\}$ under normal conditions. For online DI, we collect new measurement $\mathbf{v}^f_k$ in real time.

\subsection{Weighted moving average VMD index}
The moving average (MA) is a widely known method to improve the robustness and sensitivity. By employing a sliding time window, disturbance is filtered and detectability is improved \cite{Chen2001Principle,Ji2016Incipient,Ji2017Incipient,Chen2019Data,Chen2019Newly,Sang2020Detection}. However, the optimality of its equally weighted scheme remains unproven.
Take the least squares (LS) technique as an example. Compared with the standard LS technique, the weighted LS technique often achieves better performance when the weights are properly selected. Therefore, in this paper, we construct a weighted moving average VMD (WMA-VMD) index, which is given as follows:
\begin{align}\label{WMAVMDvf}
\widetilde{\textrm{VMD}}_{i,k}(\mathbf{a}_W,\mathbf{v}^f)=\sum\limits_{j=1}^{W}a_j\textrm{VMD}_i(\mathbf{v}^f_{k-j+1}),\;\sum\limits_{j=1}^{W}a_j=1,
\end{align}
where $\mathbf{a}_W=[a_1,a_2,\cdots,a_W]^T$ is a weight vector. For the WMA-VMD index, we put different weights on samples in the time window, as shown in \eqref{SampleTest}.
\begin{align}\label{SampleTest}
\cdots\!,\!\textrm{VMD}_i(\mathbf{v}^f_{k-W}),&\left\{\underbrace{\textrm{VMD}_i(\mathbf{v}^f_{k-W+1}),\cdots,\textrm{VMD}_i(\mathbf{v}^f_k)}\right\}\!,\!\cdots\nonumber\\
&\quad\quad\quad\{a_W,a_{W-1},\cdots,a_1\}
\end{align}
The WMA-VMD index for training data is then
\begin{align}\label{WMAVMDv}
\widetilde{\textrm{VMD}}_{i,k}(\mathbf{a}_W,\mathbf{v})=\sum\limits_{j=1}^{W}a_j\textrm{VMD}_i(\mathbf{v}_{k-j+1}).
\end{align}

In practice, $\{\textrm{VMD}_i(\mathbf{v}_k), k=1,2,\cdots,N\}$ can be viewed as a stationary process. That is, for all $k$, ${\mathbb E}\left\{\textrm{VMD}_i(\mathbf{v}_{k})\right\}=\mu_i$ and the autocorrelation function ${\mathbb C\rm{ov}}\left\{ \textrm{VMD}_i(\mathbf{v}_{k}),\textrm{VMD}_i(\mathbf{v}_{k-l}) \right\}=R_{i,l}$ depends only on the lag $l$. Then, we can derive the statistical properties of the WMA-VMD index as follows:
\begin{align}\label{StatisticalPropertyWMAVMD}
&{\mathbb E}\left\{ \widetilde{\textrm{VMD}}_{i,k}(\mathbf{a}_W,\mathbf{v}) \right\}=\sum\limits_{j=1}^{W}a_j{\mathbb E}\left\{ \textrm{VMD}_i(\mathbf{v}_{k-j+1}) \right\}=\mu_i,\nonumber\\
&{\mathbb V\rm{ar}}\left\{ \widetilde{\textrm{VMD}}_{i,k}(\mathbf{a}_W,\mathbf{v}) \right\}={\mathbb V\rm{ar}}\left\{ \sum\limits_{j=1}^{W}a_j\textrm{VMD}_i(\mathbf{v}_{k-j+1}) \right\}\nonumber\\
&=\sum\limits_{l=1}^{W}\sum\limits_{j=1}^{W}{\mathbb C\rm{ov}}\left\{ {a_l}\textrm{VMD}_i(\mathbf{v}_{k-l+1}),{a_j}\textrm{VMD}_i(\mathbf{v}_{k-j+1}) \right\}\nonumber\\
&=\sum\limits_{l=1}^{W}\sum\limits_{j=1}^{W}{a_l}{a_j}R_{i,l-j}\triangleq\tilde{S}_i(\mathbf{a}_W).
\end{align}
However, parameters $\mu_i, R_{i,l}, \tilde{S}_i(\mathbf{a}_W)$ are unknown. According to the estimation theory of the mean and the auto-covariance function of the stationary process \cite{Brockwell1991Time}, we can estimate them as follows:
\begin{align}\label{EstStatisticalPropertyWMAVMD}
&\hat{\mu}_i=\frac{1}{N}\sum\limits_{j=1}^{N}\textrm{VMD}_i(\mathbf{v}_{j}),\nonumber\\
&\hat{R}_{i,l}=\frac{1}{N}\sum\limits_{j=1}^{N-l}\left( \textrm{VMD}_i(\mathbf{v}_{j})-\hat{\mu}_i \right)\left( \textrm{VMD}_i(\mathbf{v}_{j+l})-\hat{\mu}_i \right),\nonumber\\
&\hat{R}_{i,-l}=\hat{R}_{i,l},\quad 0\leq l\leq N-1,\nonumber\\
&\hat{S}_i(\mathbf{a}_W)=\sum\limits_{l=1}^{W}\sum\limits_{j=1}^{W}{a_l}{a_j}\hat{R}_{i,l-j}.
\end{align}

\subsection{Determination of the weight}
For the WMA-VMD index, the weight vector is a crucial parameter that can directly affect the DI performance of WIOs.
It can be seen from \eqref{StatisticalPropertyWMAVMD} that the mean of WMA-VMD index is constant while its variance depends on the weight vector.
Thus, the OWV should minimize the variance. The problem formulation is given as follows.

\begin{problem}\label{PlmOptimalWeight}
For the WMA-VMD index, find the optimal weight vector $\mathbf{a}^*_W$ that minimizes its variance, i.e.,
\begin{align}
\label{EquOptFunc}
\min_{\mathbf{a}_W}\quad&\hat{S}_i(\mathbf{a}_W)=\sum\limits_{l=1}^{W}\sum\limits_{j=1}^{W}{a_l}{a_j}\hat{R}_{i,l-j},\\
\label{EquWeightSumConstraint}
{\rm s.t.}\quad& g(\mathbf{a}_W)=\sum\limits_{j=1}^{W}a_j=1.
\end{align}
\end{problem}

\begin{lemma}\label{GammaPositive}
$\hat{\mathbf{\Gamma}}^k_i, 1\leq k\leq N,$ is positive semidefinite, where
\begin{align}\label{HatGammak}
\hat{\mathbf{\Gamma}}^k_i&=\left[ {\begin{array}{cccc}
\hat{R}_{i,0} & \hat{R}_{i,-1} & \cdots & \hat{R}_{i,1-k}\\
\hat{R}_{i,1} & \hat{R}_{i,0} & \cdots & \hat{R}_{i,2-k}\\
\vdots & \vdots & \ddots & \vdots \\
\hat{R}_{i,k-1} & \hat{R}_{i,k-2} & \cdots & \hat{R}_{i,0}\end{array}} \right]\in{\mathbb R}^{k\times k}.
\end{align}
Moreover, it is positive definite if and only if $\hat{R}_{i,0}>0$.
\end{lemma}
\textbf{Proof.} Let $\epsilon_j= \textrm{VMD}_i(\mathbf{v}_{j})-\hat{\mu}_i,\ j=1,\cdots,N$. Then, we can rewrite $\hat{\mathbf{\Gamma}}^k_i=\frac{1}{N}\mathbf{\Upsilon}\mathbf{\Upsilon}^T$, where $\mathbf{\Upsilon}\in{\mathbb R}^{k\times (N+k-1)}$ and
\begin{align*}\setlength{\arraycolsep}{4.05pt}
\mathbf{\Upsilon}\!=\!\left[ {\begin{array}{ccccccccc}
0 & \cdots & 0 & \epsilon_1 & \cdots & \epsilon_{N-k+1} & \cdots & \epsilon_{N-1} &\epsilon_{N}\\
0 & \cdots & \epsilon_1 & \epsilon_2 & \cdots & \epsilon_{N-k+2} & \cdots &\epsilon_{N} &0\\
\vdots & \vdots & \vdots & \vdots & \vdots & \vdots & \vdots & \vdots & \vdots\\
\epsilon_1 & \cdots & \epsilon_{k-1} & \epsilon_{k} & \cdots & \epsilon_{N} & 0 & \cdots & 0\end{array}} \right].
\end{align*}
Thus, $\hat{\mathbf{\Gamma}}^k_i$ is positive semidefinite. Moreover, $\mathbf{\Upsilon}$ is nonsingular if and only if there is at least one nonzero $\epsilon_j$, which is equivalent to $\hat{R}_{i,0}>0$.

\begin{theorem}\label{ThmOptimalWeight}
The weight vector $\mathbf{a}^*_W$ minimizes $\hat{S}_i(\mathbf{a}_W)$ of Problem \ref{PlmOptimalWeight} is uniquely determined as
\begin{align}\label{EquOptimalWeightUnidm}
\mathbf{a}^*_W=\hat{\mathbf{A}}^{-1}\mathbf{b},
\end{align}
where $\hat{\mathbf{A}}\in{\mathbb R}^{W\times W}$, $\mathbf{b}=[0,\cdots,0,1]^T\in{\mathbb R}^{W}$,
\begin{align}\label{hatA}
\hat{A}_{l,j}=\left\{\begin{array}{ll}
\hat{R}_{i,l-j}-\hat{R}_{i,l+1-j},&{}\quad l<W,\\
1,&{}\quad l=W.
\end{array}\right.
\end{align}
\end{theorem}
\textbf{Proof.} Since this is a constrained optimization problem, we construct a Lagrange function as follows:
\begin{align}
{\cal L}(\mathbf{a}_W,\lambda)=\frac{1}{2}\sum\limits_{l=1}^{W}\sum\limits_{j=1}^{W}{a_l}{a_j}\hat{R}_{i,l-j}+\lambda(\sum\limits_{j=1}^{W}a_j-1),
\end{align}
where $\lambda$ is a Lagrange multiplier. Note that
\begin{align*}
&\frac{\partial{\cal L}(\mathbf{a}_W,\lambda)}{\partial a_l}=\sum\limits_{j=1}^{W}a_j\hat{R}_{i,l-j}+\lambda,\\
&\nabla^2_{\mathbf{a}_W}\hat{S}_i(\mathbf{a}_W)=2\hat{\mathbf{\Gamma}}^W_i\succ \mathbf{0},
\end{align*}
where the last inequality is because of \textit{Lemma \ref{GammaPositive}}. Thus, \textit{Problem \ref{PlmOptimalWeight}} is a convex optimization.
Then, a weight vector $\mathbf{a}^*_W$ minimizes $\hat{S}_i(\mathbf{a}_W)$ of \textit{Problem \ref{PlmOptimalWeight}} if and only if it satisfies the Karush-Kuhn-Tucker conditions \cite{Luenberger2008Linear}, i.e., $\nabla_{\mathbf{a}_W}{\cal L}(\mathbf{a}_W,\lambda)=\mathbf{0}_W,\ \nabla_{\lambda}{\cal L}(\mathbf{a}_W,\lambda)=0$.
By setting the derivative of ${\cal L}(\mathbf{a}_W,\lambda)$ with respect to $\mathbf{a}_W$ to zeros, we obtain
\begin{align}
\label{DerivativeEqual}
\sum\limits_{j=1}^{W}a_j(\hat{R}_{i,l-j}-\hat{R}_{i,l'-j})=0,\ 1\leq l,l'\leq W.
\end{align}
Integrating \eqref{DerivativeEqual} with \eqref{EquWeightSumConstraint}, we derive $\hat{\mathbf{A}}\mathbf{a}^*_W=\mathbf{b}$.

Now, we prove that $\hat{\mathbf{A}}$ is nonsingular so that the OWV is unique.
By following a few reformulations, we can rewrite $\hat{\mathbf{A}}=\hat{\mathbf{J}}\hat{\mathbf{\Gamma}}^{W}_i$, where
\begin{align*}
\hat{\mathbf{J}}=\left[ \begin{array}{c}
\underline{\begin{array}{ccccc}
1 & -1 & 0 & \cdots & 0\\
0 & 1 & -1 & \ddots & \vdots\\
\vdots & \ddots & \ddots & \ddots & 0\\
0 & \cdots & 0 & 1 & -1\end{array}} \\
\mathbf{1}^T_W(\hat{\mathbf{\Gamma}}^{W}_i)^{-1}
\end{array}\right]\in{\mathbb R}^{W\times W}.
\end{align*}
For $|\hat{\mathbf{J}}|$, adding its $j$th column to its $j\!-\!1$th column in turn, we obtain $|\hat{\mathbf{J}}|=\mathbf{1}^T_W(\hat{\mathbf{\Gamma}}^{W}_i)^{-1}\mathbf{1}_W>0$.
Thus, $\hat{\mathbf{A}}$ is nonsingular and the proof is complete.

\subsection{Properties of the optimal weight vector}
\begin{theorem}\label{ThmSymmetryAW}
(Symmetry of the OWV) The OWV $\mathbf{a}^*_W=[a^*_1,a^*_2,\cdots,a^*_W]^T$ for Problem \ref{PlmOptimalWeight} satisfies
\begin{align}\label{EquSymmetryAW}
a^*_m=a^*_{W-m+1},\quad 1\leq m\leq W.
\end{align}
\end{theorem}
\textbf{Proof.} It can be seen from \eqref{hatA} that
\begin{align*}
\begin{array}{ll}
\hat{A}_{l,j}=\hat{A}_{l+1,j+1},&{}\quad l<W-1, j<W,\\
\hat{A}_{l,j}=-\hat{A}_{j-1,l},&{}\quad l\leq W-1, j\leq W.
\end{array}
\end{align*}
Denote $\hat{\mathbf{A}}_{\backslash l\backslash\emptyset}$ and $\hat{\mathbf{A}}_{\backslash\emptyset\backslash j}$ as the matrices obtained from $\hat{\mathbf{A}}$ by deleting the $l$th row and the $j$th column, respectively. Then, $\hat{\mathbf{A}}_{\backslash W\backslash\emptyset}\in{\mathbb R}^{(W-1)\times W}$ has the following form:
\begin{align*}
\hat{\mathbf{A}}_{\backslash W\backslash\emptyset}&=\left[ {\begin{array}{cccccc}
-t_1 & t_1 & t_2 & \cdots & t_{W-2} & t_{W-1}\\
-t_2 & -t_1 & t_1 & t_2 & \ddots & t_{W-2}\\
-t_3 & -t_2 & -t_1 & t_1 & \ddots & \vdots\\
\vdots & \ddots & \ddots & \ddots & \ddots & t_2\\
-t_{W-1} & \cdots & -t_3 & -t_2 & -t_1 & t_1\\
\end{array}} \right],
\end{align*}
where $t_l=\hat{R}_{i,l}-\hat{R}_{i,l-1}$.
Define $\textrm{cen}(\mathbf{A})\in{\mathbb R}^{p\times q}$ as the centrosymmetry of a matrix $\mathbf{A}\in{\mathbb R}^{p\times q}$, namely, $[\textrm{cen}(\mathbf{A})]_{l,j}\!=\![\mathbf{A}]_{p-l+1,q-j+1}$.
It can be easily verified that
\begin{align}\label{CenProperty1}
\textrm{cen}\left(\textrm{cen}(\mathbf{A})\right)=\mathbf{A},\quad\textrm{cen}(-\mathbf{A})=-\textrm{cen}(\mathbf{A}).
\end{align}
Besides, if $\mathbf{A}$ is a square matrix, then we have $|\textrm{cen}(\mathbf{A})|=|\mathbf{A}|$.
Moreover, if $\mathbf{A}$ is centrosymmetric, that is to say, $\textrm{cen}(\mathbf{A})=\mathbf{A}$, we have
\begin{align}\label{CenProperty2}
\textrm{cen}(\mathbf{A}_{\backslash\emptyset\backslash j})=\mathbf{A}_{\backslash\emptyset\backslash q-j+1}.
\end{align}
As for $\hat{\mathbf{A}}$, note that $[\hat{\mathbf{A}}_{\backslash W\backslash\emptyset}]_{l,j}\!=\!-[\hat{\mathbf{A}}_{\backslash W\backslash\emptyset}]_{W-l,W-j+1}$, i.e., $\textrm{cen}(\hat{\mathbf{A}}_{\backslash W\backslash\emptyset})\!=\!-\hat{\mathbf{A}}_{\backslash W\backslash\emptyset}$. According to \eqref{CenProperty1} and \eqref{CenProperty2}, we have
\begin{align}
\textrm{cen}(\hat{\mathbf{A}}_{\backslash W\backslash m})=-\hat{\mathbf{A}}_{\backslash W\backslash W-m+1}.
\end{align}
Thus,
\begin{align*}
|\hat{\mathbf{A}}_{\backslash W\backslash m}|&=|-\textrm{cen}(\hat{\mathbf{A}}_{\backslash W\backslash W-m+1})|\\
&=(-1)^{W-1}|\hat{\mathbf{A}}_{\backslash W\backslash W-m+1}|.
\end{align*}
Note that $\mathbf{a}^*_W=\hat{\mathbf{A}}^{-1}\mathbf{b}=|\hat{\mathbf{A}}|^{-1}\textrm{adj}(\hat{\mathbf{A}})\mathbf{b}$. Then,
\begin{align}\label{amExpansion}
a^*_m&=(-1)^{W+m}|\hat{\mathbf{A}}|^{-1}\left|\hat{\mathbf{A}}_{\backslash W\backslash m}\right|\\
&=(-1)^{2W-m+1}|\hat{\mathbf{A}}|^{-1}\left|\hat{\mathbf{A}}_{\backslash W\backslash W-m+1}\right|=a^*_{W-m+1},\nonumber
\end{align}
which completes the proof.

\begin{corollary}
For $W=2$, the OWV $\mathbf{a}^*_W$ minimizes $\hat{S}_i(\mathbf{a}_W)$ of Problem \ref{PlmOptimalWeight} is uniquely determined as $a^*_1=a^*_2=1/2$.
\end{corollary}
\textbf{Proof.} Directly derived from Theorems \ref{ThmOptimalWeight} and \ref{ThmSymmetryAW}.

\begin{remark}
Intuitively, since the process is stationary, the first and last samples in a time window always have the same contributions to the covariance matrices $\tilde{S}_i(\mathbf{a}_W)$, as well as $\hat{S}_i(\mathbf{a}_W)$, as can be seen in \eqref{StatisticalPropertyWMAVMD} and \eqref{EstStatisticalPropertyWMAVMD}. Therefore, they should have the same weight. This is also true for the second and the penultimate samples, and so on. \textit{Theorem \ref{ThmSymmetryAW}} reveals that the OWV possesses a symmetrical structure, and helps us better understand the WMA technique.
\end{remark}

\begin{theorem}\label{ThmOptimalWeightIndp}
(Optimality of the equally weighted scheme for independent data) When the process data are independent, the weight vector $\mathbf{a}^*_W$ minimizes $\hat{S}_i(\mathbf{a}_W)$ of Problem \ref{PlmOptimalWeight} is uniquely determined as
\begin{align}\label{EquOptimalWeightIndp}
a^*_1=a^*_2=\cdots=a^*_W=1/W.
\end{align}
\end{theorem}
\textbf{Proof.} When the process data are independent, we have $\hat{R}_{i,l}=0$, $\forall l\neq 0$. Thus, in this case, we have
\begin{align*}
\hat{\mathbf{A}}=\left[ \begin{array}{c}
\begin{array}{ccccc}
\hat{R}_{i,0} & -\hat{R}_{i,0} & 0 & \cdots & 0\\
0 & R_{i,0} & -\hat{R}_{i,0} & \ddots & \vdots\\
\vdots & \ddots & \ddots & \ddots & 0\\
0 & \cdots & 0 & \hat{R}_{i,0} & -\hat{R}_{i,0}\\
1 & \cdots & 1 & \cdots & 1\end{array} \\
\end{array}\right],
\end{align*}
and
\begin{align*}
|\hat{\mathbf{A}}|=W(\hat{R}_{i,0})^{W-1},\ |\hat{\mathbf{A}}_{\backslash W\backslash m}|=(-1)^{W-m}(\hat{R}_{i,0})^{W-1}.
\end{align*}
According to \eqref{amExpansion}, $a^*_m=(-1)^{W+m}|\hat{\mathbf{A}}|^{-1}\left|\hat{\mathbf{A}}_{\backslash W\backslash m}\right|=1/W$.

\begin{remark}
\textit{Theorem \ref{ThmOptimalWeightIndp}} proves the optimality of the MA technique's equally weighted scheme when data are independent. It explains why the MA technique is always adopted in system monitoring tasks where samples are assumed to be independent, such as in \cite{Chen2001Principle,Ji2016Incipient,Ji2017Incipient,Sang2020Detection}.
\end{remark}

\begin{theorem}\label{ThmOptimalWeightPositive}
For the OWV $\mathbf{a}^*_W=[a^*_1,a^*_2,\cdots,a^*_W]^T$ minimizing $\hat{S}_i(\mathbf{a}_W)$ of Problem \ref{PlmOptimalWeight}, we have $a^*_m>(=,<)0$ if and only if $|\check{\mathbf{\Gamma}}^{W}_i(m)|>(=,<)0$, where
\begin{align}\label{EquOptimalWeightPositive}
[\check{\Gamma}^{W}_i(m)]_{l,j}=\left\{\begin{array}{ll}
[\hat{\Gamma}^{W}_i]_{l,j},&{}\quad j\neq m,\vspace{0.1cm}\\
1,&{}\quad j=m.
\end{array}\right.
\end{align}
\end{theorem}
\textbf{Proof.} According to \eqref{EquOptimalWeightUnidm} and the Cramer's rule, we have
\begin{align}
a^*_m=|\hat{\mathbf{A}}|^{-1}|\hat{\mathbf{B}}^{m}|,
\end{align}
where $\hat{\mathbf{B}}^{m}$ is the matrix obtained by replacing the $m$th column of $\hat{\mathbf{A}}$ by $\mathbf{b}$. By following the properties of determinants, we have $|\hat{\mathbf{B}}^{m}|\!=\!|\check{\mathbf{A}}^m|$, where $\delta_{mj}$ is the Kronecker function and
\begin{align}\label{checkAm}
\check{A}^m_{l,j}=\left\{\begin{array}{ll}
\hat{A}_{l,j},&{}\quad l<W,\vspace{0.1cm}\\
\delta_{mj},&{}\quad l=W.
\end{array}\right.
\end{align}
Moreover, by following a few reformulations, we can rewrite $\check{\mathbf{A}}^m=\check{\mathbf{J}}^m\hat{\mathbf{\Gamma}}^{W}_i$, where
\begin{align*}
\check{\mathbf{J}}^m=\left[ \begin{array}{c}
\underline{\begin{array}{ccccc}
1 & -1 & 0 & \cdots & 0\\
0 & 1 & -1 & \ddots & \vdots\\
\vdots & \ddots & \ddots & \ddots & 0\\
0 & \cdots & 0 & 1 & -1\end{array}} \\
\mathbf{e}^T_{Wm}(\hat{\mathbf{\Gamma}}^{W}_i)^{-1}
\end{array}\right]\in{\mathbb R}^{W\times W}.
\end{align*}
For $|\check{\mathbf{J}}^m|$, adding its $j$th column to its $j\!-\!1$th column in turn, we obtain $|\check{\mathbf{J}}^m|=\mathbf{e}^T_{Wm}(\hat{\mathbf{\Gamma}}^{W}_i)^{-1}\mathbf{1}_W$. Then,
\begin{align}
a^*_m=|\hat{\mathbf{A}}|^{-1}|\check{\mathbf{A}}^m|=|\hat{\mathbf{J}}|^{-1}|\check{\mathbf{J}}^m|.
\end{align}
Recall that $|\hat{\mathbf{J}}|=\mathbf{1}^T_W(\hat{\mathbf{\Gamma}}^{W}_i)^{-1}\mathbf{1}_W>0$. Thus, we have $a^*_m>(=,<)0$ if and only if $|\check{\mathbf{J}}^m|=\mathbf{e}^T_{Wm}(\hat{\mathbf{\Gamma}}^{W}_i)^{-1}\mathbf{1}_W>(=,<)0$, which is further equivalent to
\begin{align}
\mathbf{e}^T_{Wm}\hat{\mathbf{M}}^{W}_i\mathbf{1}_W>(=,<)0,
\end{align}
where $\hat{\mathbf{M}}^{W}_i$ is the adjoint of $\hat{\mathbf{\Gamma}}^{W}_i$. Note that $[\hat{M}^{W}_i]_{l,j}$ is the algebraic cofactor of $[\hat{\Gamma}^{W}_i]_{j,l}$. Thus, we have
\begin{align}
\mathbf{e}^T_{Wm}\hat{\mathbf{M}}^{W}_i\mathbf{1}_W=\sum\limits_{j=1}^{W}[\hat{M}^{W}_i]_{m,j}=|\check{\mathbf{\Gamma}}^{W}_i(m)|,
\end{align}
which completes the proof.

\begin{remark}
\textit{Theorem \ref{ThmOptimalWeightPositive}} gives a necessary and sufficient condition for the OWV being positive, which will help us analyze the WMA-VMD-based detectability and isolability of IOs in Section \ref{DIanalysisWMAVMDSub}.
\end{remark}

\subsection{Detection and isolation using WMA-VMD}
In traction mode, the $i$th wheelset's creep is considered normal at time instance $k$, if $\widetilde{\textrm{VMD}}_{i,k}(\mathbf{a}^{*}_W,\mathbf{v}^f)\leq\delta_{i,W}$. Otherwise, the $i$th wheelset is considered to be slipping. The control limit (CL) $\delta_{i,W}$ can be determined based on the historical data as follows:
\begin{align}\label{PWMAVMDCL}
\delta_{i,W}=\max_{k\in\{W,W+1,\cdots,N\}}\widetilde{\textrm{VMD}}_{i,k}(\mathbf{a}^*_W,\mathbf{v}).
\end{align}

In braking mode, the procedure is the same except that we should use $-\mathbf{v},-\mathbf{v}^f$ instead of $\mathbf{v},\mathbf{v}^f$ in all of the above steps. Overall, the DI algorithm of WIOs based on the WMA-VMD index with window length $W$ is summarized as Algorithm 1.
\begin{table}\centering
\begin{tabular}{p{8cm}}\hline
{\bf Algorithm 1:} DI of the $i$th wheelset's IOs\\\hline
{\bf Initialization:} Collect training samples $\{\mathbf{v}_k, k\!=\!1,\cdots,N\}$ under normal conditions, and test samples $\{\mathbf{v}^f_{k-W+1},\cdots,\mathbf{v}^f_k\}$ in real time. For traction mode, set $M=1$. For braking mode, set $M=-1$.\\
\end{tabular}
\begin{tabular}{p{0.1cm}p{7.9cm}}
   {}& {\bf Off-line Calculation:}\\
   1.& Set $\mathbf{v}_k=M\times\mathbf{v}_k,\ k=1,2,\cdots,N$.\\
   2.& Compute $\textrm{VMD}_i(\mathbf{v})$ by \eqref{VMD} and $\hat{R}_{i,l}$ by \eqref{EstStatisticalPropertyWMAVMD}.\\
   3.& Calculate the OWV $\mathbf{a}^*_W$ by \eqref{EquOptimalWeightUnidm}{\dag}.\\
   4.& Determine the CL $\delta_{i,W}$ by \eqref{PWMAVMDCL}{\dag}.\\
   {}& {\bf On-line detection and isolation:}\\
   1.& Set $\mathbf{v}^f_j=M\times\mathbf{v}^f_j,\ j=k\!-\!W\!+\!1,\cdots,k$.\\
   2.& Compute the real-time WMA-VMD index $\widetilde{\textrm{VMD}}_{i,k}(\mathbf{a}^{*}_W,\mathbf{v}^f)$ by \eqref{WMAVMDvf}.\\
   3.& When $\widetilde{\textrm{VMD}}_{i,k}(\mathbf{a}^{*}_W,\mathbf{v}^f)>\delta_{i,W}$, the $i$th wheelset is considered to be slipping if $M\!=\!1$, or sliding if $M\!=\!-1$.\\\hline
\end{tabular}
\begin{tablenotes}
\item{\dag} $\mathbf{a}^*_W$, as well as $\delta_{i,W}$, can be not the same for different $i$ or $M$.
\end{tablenotes}
\end{table}

%The WMA-VMD statistic with window length $W$ to detect and isolate the $i$th wheelset's IOs, denoted as WMA-$\textrm{VMD}_i(W)$ at time instance $k$ is then
%\begin{align}
%\textrm{WMA-VMD}^{f}_{i,k}(W)&=\widetilde{\textrm{VMD}}_{i,k}(\mathbf{a}^{*}_W,\mathbf{v}^f).
%\end{align}
%as pointed out in Section \ref{VMDindex}, we should use $-\mathbf{v}$ to determine the optimal weight $\mathbf{a}^{*}_W$ and calculate the WMA-$\textrm{VMD}_i$ statistic.
%where the WMA-$\textrm{VMD}_i$ statistic is calculated using $-\mathbf{v}^f_k$ instead of $\mathbf{v}^f_k$.
%First transfer the sample $\mathbf{v}$ to $\mathbf{v}'=-\mathbf{v}$ and then use the $\mathbf{v}'$ to determine the optimal weight by following (41-43) and \textit{Theorem \ref{ThmOptimalWeight}}. Denote the optimal weight using $\mathbf{v}$ as

\section{Detectability and isolability analysis}\label{DIanalysisSec}
In this section, we analyze the WMA-VMD-based detectability and isolability of WIOs. To this end, the properties of the VMD operator are first derived. Then, necessary conditions and sufficient conditions for WMA-VMD-based detectability and isolability of WIOs are given.

\subsection{Properties of the VMD operator}
In this subsection, important properties, such as the triangle inequality, of the VMD operator are proven, see \textit{Theorem \ref{ThmVMDProperty}}. We begin with the proof of some properties of the $\min$ operator.

\begin{lemma}\label{LemScalarMin}
For any scalars $a,b,c,d\in{\mathbb R}^{1}$, we have
\begin{align}
\label{EquScalarMin1}
\min(\min(a,b),c)&=\min(a,b,c),\\
\label{EquScalarMin2}
\min(a\!+\!c,a\!+\!d)&=a+\min(c,d),\\
\label{EquScalarMin3}
\min(a,b)+\min(c,d)&\leq\min(a\!+\!c,b\!+\!d),\\
\label{EquScalarMin4}
\min(a\!-\!c,b\!-\!d)&\leq\min(a,b)-\min(c,d).
\end{align}
Furthermore, the equality in \eqref{EquScalarMin3} holds if and only if
\begin{align}\label{EquScalarMin3Equ}
(\mathrm{i})\ a=b;\quad\qquad &\mathrm{or}\ (\mathrm{ii})\ c=d;\nonumber\\
\mathrm{or}\ (\mathrm{iii})\ a<b, c<d;\quad&\mathrm{or}\ (\mathrm{iv})\ b<a, d<c.
\end{align}
The equality in \eqref{EquScalarMin4} holds if and only if
\begin{align}\label{EquScalarMin4Equ}
(\mathrm{i})\ c=d;&\qquad\mathrm{or}\ (\mathrm{ii})\ a\leq b, c<d, a-c\leq b-d;\nonumber\\
&\mathrm{or}\ (\mathrm{iii})\ b\leq a, d<c, b-d\leq a-c.
\end{align}
\end{lemma}
\textbf{Proof.} Equalities \eqref{EquScalarMin1} and \eqref{EquScalarMin2} are obvious. As for inequality \eqref{EquScalarMin3}, its left side has four possible values, i.e.,
\begin{align*}
\min(a,b)+\min(c,d)=\left\{\begin{array}{ll}
a+c,&{}\quad a\leq b, c\leq d,\\
a+d,&{}\quad a\leq b, d\leq c,\\
b+c,&{}\quad b\leq a, c\leq d,\\
b+d,&{}\quad b\leq a, d\leq c.
\end{array}\right.
\end{align*}
As for the right side of inequality \eqref{EquScalarMin3}, we have
\begin{align*}
\min(a\!+\!c,b\!+\!d)=\left\{\begin{array}{ll}
a+c,&{}\quad a\leq b, c\leq d,\\
b+d,&{}\quad b\leq a, d\leq c.
\end{array}\right.
\end{align*}
Moreover, when $a\leq b, d\leq c$, the right side of inequality \eqref{EquScalarMin3} has two possible values, i.e.,
\begin{align*}
\min(a\!+\!c,b\!+\!d)=\left\{\begin{array}{ll}
a+c,&{}\quad a+c\leq b+d,\\
b+d,&{}\quad b+d\leq a+c.
\end{array}\right.
\end{align*}
It can be easily seen that both of them are no less than $a+d$ since $a\leq b, d\leq c$.
Similarly, when $b\leq a, c\leq d$, the right side of inequality \eqref{EquScalarMin3} also has these two possible values, both of which are no less than $b+c$ since $b\leq a, c\leq d$.
Based on the above discussions, we obtain \eqref{EquScalarMin3}.
Furthermore, following the above proof, we can conclude that the equality in \eqref{EquScalarMin3} holds if and only if at least one of the following conditions is satisfied:
\begin{align*}
(\mathrm{i})\ a\leq b, c\leq d; \qquad (\mathrm{ii})\ b\leq a, d\leq c;&\\
(\mathrm{iii})\ a\leq b, d\leq c, a+c\leq b+d, c=d;&\\
(\mathrm{iv})\ a\leq b, d\leq c, b+d\leq a+c, a=b;&\\
(\mathrm{v})\ b\leq a, c\leq d, a+c\leq b+d, a=b;&\\
(\mathrm{vi})\ b\leq a, c\leq d, b+d\leq a+c, c=d.&
\end{align*}
The above conditions can be simplified as
\begin{align*}
(\mathrm{i})\ a\leq b, c\leq d; \qquad (\mathrm{ii})\ b\leq a, d\leq c;&\\
(\mathrm{iii})\ a\leq b, c=d; \qquad (\mathrm{iv})\ a=b, d\leq c;&\\
(\mathrm{v})\ a=b, c\leq d; \qquad (\mathrm{vi})\ b\leq a, c=d;&
\end{align*}
which is further equivalent to \eqref{EquScalarMin3Equ}.

As for inequality \eqref{EquScalarMin4}, its right side has four possible values, i.e.,
\begin{align*}
\min(a,b)-\min(c,d)=\left\{\begin{array}{ll}
a-c,&{}\quad a\leq b, c\leq d,\\
a-d,&{}\quad a\leq b, d\leq c,\\
b-c,&{}\quad b\leq a, c\leq d,\\
b-d,&{}\quad b\leq a, d\leq c.
\end{array}\right.
\end{align*}
As for the left side of inequality \eqref{EquScalarMin4}, we have
\begin{align*}
\min(a\!-\!c,b\!-\!d)=\left\{\begin{array}{ll}
a-c,&{}\quad a\leq b, d\leq c,\\
b-d,&{}\quad b\leq a, c\leq d.
\end{array}\right.
\end{align*}
It can be seen that \eqref{EquScalarMin4} holds in these two cases.
Moreover, when $a\leq b, c\leq d$ or $b\leq a, d\leq c$, \eqref{EquScalarMin4} also holds since
\begin{align*}
\min(a\!-\!c,b\!-\!d)\leq a-c,\quad\min(a\!-\!c,b\!-\!d)\leq b-d.
\end{align*}
Based on the above discussions, we obtain \eqref{EquScalarMin4}.
Furthermore, following the above proof, we can conclude that the equality in \eqref{EquScalarMin4} holds if and only if at least one of the following conditions is satisfied:
\begin{align*}
&(\mathrm{i})\,a\leq b, c\leq d, a\!-\!c\leq b\!-\!d; (\mathrm{ii})\,a\leq b, c\leq d, a\!-\!c=b\!-\!d;\\
&(\mathrm{iii})\,a\leq b, d\leq c, c=d; \quad(\mathrm{iv})\,b\leq a, c\leq d, c=d;\\
&(\mathrm{v})\,b\leq a, d\leq c, b\!-\!d=a\!-\!c; (\mathrm{vi})\,b\leq a, d\leq c, b\!-\!d\leq a\!-\!c;
\end{align*}
which is further equivalent to \eqref{EquScalarMin4Equ}. The proof is now complete.
%(i) $a\leq b, c\leq d$; (ii) $b\leq a, d\leq c$; (iii) $a\leq b, d\leq c, a+c\leq b+d, c=d$; (iv) $a\leq b, d\leq c, b+d\leq a+c, a=b$; (v) $b\leq a, c\leq d, a+c\leq b+d, a=b$; (vi) $b\leq a, c\leq d, b+d\leq a+c, c=d$.
%(i) $a\leq b, c\leq d$; (ii) $b\leq a, d\leq c$; (iii) $a\leq b, c=d$; (iv) $a=b, d\leq c$; (v) $a=b, c\leq d$; (vi) $b\leq a, c=d$.
%(i) $a=b$, or (ii) $c=d$, or (iii) $a<b, c<d$, or (iv) $b<a, d<c$.

\begin{lemma}\label{LemMinProperty}
Given $\mathbf{x},\mathbf{y}\in{\mathbb R}^{p}$, $z\in{\mathbb R}^{1}$, then
\begin{align}
\label{EquMinProperty1}
&\min(z\mathbf{x})=\left\{\begin{array}{ll}
z\min(\mathbf{x}),&{}\quad z\geq 0,\\
z\max(\mathbf{x}),&{}\quad z<0,
\end{array}\right.\\
\label{EquMinProperty2}
&\min(\mathbf{x})+\min(\mathbf{y})\leq\min(\mathbf{x}+\mathbf{y}),\\
\label{EquMinProperty3}
&\min(\mathbf{x})-\min(\mathbf{y})\geq\min(\mathbf{x}-\mathbf{y}),
\end{align}
where $\min(\mathbf{x})\!=\!\min(x_{1},\!\cdots\!,x_p),\max(\mathbf{x})\!=\!\max(x_{1},\!\cdots\!,x_p)$.
\end{lemma}
\textbf{Proof.} When $z\geq 0$, equality \eqref{EquMinProperty1} is obvious. When $z<0$,
\begin{align*}
\min(z\mathbf{x})=\min(-\|z\|\mathbf{x})=\|z\|\min(-\mathbf{x})=z\max(\mathbf{x}),
\end{align*}
where the last equality is because of \eqref{MinConvertMax}. Let $\mathbf{x}^i\!=\![x_1, x_2,\cdots,x_i]^T,\mathbf{y}^i\!=\![y_1, y_2,\cdots,y_i]^T\in{\mathbb R}^{i}$. According to \textit{Lemma \ref{LemScalarMin}}, we have
\begin{align}
\label{EquMinProperty1P1}
\min&(\mathbf{x}^p)+\min(\mathbf{y}^p)\nonumber\\
&=\min\left( \min(\mathbf{x}^{p-1}),x_p \right)+\min\left( \min(\mathbf{y}^{p-1}),y_p \right)\nonumber\\
&\leq\min\left( \min(\mathbf{x}^{p-1})+\min(\mathbf{y}^{p-1}),x_p+y_p \right),\\
\label{EquMinProperty2P1}
\min&(\mathbf{x}^p)-\min(\mathbf{y}^p)\nonumber\\
&=\min\left( \min(\mathbf{x}^{p-1}),x_p \right)-\min\left( \min(\mathbf{y}^{p-1}),y_p \right)\nonumber\\
&\geq\min\left( \min(\mathbf{x}^{p-1})-\min(\mathbf{y}^{p-1}),x_p-y_p \right).
\end{align}
Likewise, we have
\begin{align}
\label{EquMinProperty1P2}
\min&(\mathbf{x}^{p-1})+\min(\mathbf{y}^{p-1})\nonumber\\
&\leq\min\left( \min(\mathbf{x}^{p-2})+\min(\mathbf{y}^{p-2}),x_{p-1}+y_{p-1} \right),\\
\label{EquMinProperty2P2}
\min&(\mathbf{x}^{p-1})-\min(\mathbf{y}^{p-1})\nonumber\\
&\geq\min\left( \min(\mathbf{x}^{p-2})-\min(\mathbf{y}^{p-2}),x_{p-1}+y_{p-1} \right).
\end{align}
Substituting \eqref{EquMinProperty1P2} into \eqref{EquMinProperty1P1}, and \eqref{EquMinProperty2P2} into \eqref{EquMinProperty2P1} respectively, we have
\begin{align*}
&\min(\mathbf{x}^p)+\min(\mathbf{y}^p)\nonumber\\
&\leq\min\left( \min\left( \min(\mathbf{x}^{p-2})+\min(\mathbf{y}^{p-2}),x_{p-1}\!+\!y_{p-1} \right),x_p\!+\!y_p \right)\nonumber\\
&=\min\left( \min(\mathbf{x}^{p-2})+\min(\mathbf{y}^{p-2}),x_{p-1}\!+\!y_{p-1},x_p\!+\!y_p \right),\\
&\min(\mathbf{x}^p)-\min(\mathbf{y}^p)\nonumber\\
&\geq\min\left( \min\left( \min(\mathbf{x}^{p-2})-\min(\mathbf{y}^{p-2}),x_{p-1}\!-\!y_{p-1} \right),x_p\!-\!y_p \right)\nonumber\\
&=\min\left( \min(\mathbf{x}^{p-2})-\min(\mathbf{y}^{p-2}),x_{p-1}\!-\!y_{p-1},x_p\!-\!y_p \right).
\end{align*}
Continuing the recursion above, we obtain
\begin{align*}
&\min(\mathbf{x})+\min(\mathbf{y})=\min(\mathbf{x}^p)+\min(\mathbf{y}^p)\nonumber\\
&\leq\min\left( x_{1}+y_{1},\cdots,x_{p-1}+y_{p-1},x_p+y_p \right)=\min(\mathbf{x}+\mathbf{y}),\\
&\min(\mathbf{x})-\min(\mathbf{y})=\min(\mathbf{x}^p)-\min(\mathbf{y}^p)\nonumber\\
&\geq\min\left( x_{1}-y_{1},\cdots,x_{p-1}-y_{p-1},x_p-y_p \right)=\min(\mathbf{x}-\mathbf{y}),
\end{align*}
which completes the proof.

\begin{theorem}\label{ThmVMDProperty}
Given $\mathbf{x},\mathbf{y}\in{\mathbb R}^{p}$, $z\in{\mathbb R}^{1}$ and $i\in\{1,2,\cdots,p\}$, then
\begin{align}
\label{EquVMDProperty1}
\textrm{VMD}_i(\mathbf{x})=0\ &\Leftrightarrow\ x_i=\min(\mathbf{x}),\\
\label{EquVMDProperty2}
\textrm{VMD}_i(\mathbf{x}+z\mathbf{1}_p)&=\textrm{VMD}_i(\mathbf{x}),\\
\label{EquVMDProperty3}
\textrm{VMD}_i(z\mathbf{x})&=\left\{\begin{array}{ll}
z\textrm{VMD}_i(\mathbf{x}),&{}\ z\geq 0,\\
-z\textrm{VMD}_i(-\mathbf{x}),&{}\ z<0,
\end{array}\right.\\
\label{EquVMDProperty4}
\textrm{VMD}_i(\mathbf{x}+\mathbf{y})&\leq\textrm{VMD}_i(\mathbf{x})+\textrm{VMD}_i(\mathbf{y}),\\
\label{EquVMDProperty5}
\textrm{VMD}_i(\mathbf{x}-\mathbf{y})&\geq\textrm{VMD}_i(\mathbf{x})-\textrm{VMD}_i(\mathbf{y}).
\end{align}
\end{theorem}
\textbf{Proof.} Equality \eqref{EquVMDProperty1} is obvious. Equality \eqref{EquVMDProperty2} can be derived from \eqref{EquScalarMin1} and \eqref{EquScalarMin2} directly, and \eqref{EquVMDProperty3} can be derived from \eqref{EquMinProperty1}. According to \textit{Lemma \ref{LemMinProperty}}, we have
\begin{align*}
&\textrm{VMD}_i(\mathbf{x}+\mathbf{y})=x_i+y_i-\min(\mathbf{x}+\mathbf{y})\\
&\leq x_i+y_i-\min(\mathbf{x})-\min(\mathbf{y})=\textrm{VMD}_i(\mathbf{x})+\textrm{VMD}_i(\mathbf{y}),\\
&\textrm{VMD}_i(\mathbf{x}-\mathbf{y})=x_i-y_i-\min(\mathbf{x}-\mathbf{y})\\
&\geq x_i-y_i-\min(\mathbf{x})+\min(\mathbf{y})=\textrm{VMD}_i(\mathbf{x})-\textrm{VMD}_i(\mathbf{y}),
\end{align*}
which completes the proof.

\subsection{Analysis for the VMD index}\label{DIanalysisVMDSub}
Velocity measurements with WIOs can be described as
\begin{align}\label{IOModel}
\mathbf{v}^f_k=\mathbf{v}^*_k+\Xi_k\mathbf{f}_k,
\end{align}
where $\mathbf{v}^*_k$ denotes the normal velocity fluctuation, $\Xi_k$ is the direction matrix of WIOs in time instance $k$, and $\mathbf{f}_k$ is the WIOs' magnitude vector.
Model \eqref{IOModel} can represent several kinds of WIOs. For example, a WIO occurring on the second wheelset can be described by the direction matrix
\begin{align}
\Xi=[0, 1, 0, \cdots]^T\in{\mathbb R}^{p}.
\end{align}
When a WIO occurs on both the first and second wheelsets, the corresponding IO direction matrix is
\begin{align}
\Xi&=\left[ {\begin{array}{cccc}
1 & 0 & 0 & \cdots\\
0 & 1 & 0 & \cdots\end{array}} \right]^T\in{\mathbb R}^{p\times 2}.
\end{align}
Moreover, since $\mathbf{v}^*_k$ denotes the normal velocity fluctuation, i.e., neither slip nor slide occurring, it satisfies
\begin{align}\label{VMDNorConstraints}
\textrm{VMD}_i(\mathbf{v}^*_k)\leq\delta_{i,1},\ \textrm{VMD}_i(-\mathbf{v}^*_k)\leq\phi_{i,1},
\end{align}
where $\delta_{i,1}$ and $\phi_{i,1}$ are calculated by \eqref{PWMAVMDCL} in traction and braking mode respectively, and are given by
\begin{align}
\label{PVMDCL}
\delta_{i,1}&=\max_{k\in\{1,2,\cdots,N\}}\textrm{VMD}_i(\mathbf{v}_k),\\
\label{NVMDCL}
\phi_{i,1}&=\max_{k\in\{1,2,\cdots,N\}}\textrm{VMD}_i(-\mathbf{v}_k).
\end{align}

\begin{theorem}\label{ThmVMDNIC}
For the VMD index, a necessary isolability condition of the $i$th wheelset's IO is
\begin{align}\label{EquVMDNIC}
\textrm{VMD}_i(\Xi_k\mathbf{f}_k)\neq 0.
\end{align}
\end{theorem}
\textbf{Proof.} Substitute the WIO model \eqref{IOModel} into the VMD index. Then, following properties of the $\textrm{VMD}_i()$ operator, we have
\begin{align}\label{VfVMDLeq}
\textrm{VMD}_i(\mathbf{v}^f_k)&=\textrm{VMD}_i(\mathbf{v}^*_k+\Xi_k\mathbf{f}_k)\nonumber\\
&\leq\textrm{VMD}_i(\mathbf{v}^*_k)+\textrm{VMD}_i(\Xi_k\mathbf{f}_k).
\end{align}
If $\textrm{VMD}_i(\Xi_k\mathbf{f}_k)=0$, the $i$th wheelset's IO is not isolable. Because in this case, we have
\begin{align*}
\textrm{VMD}_i(\mathbf{v}^f_k)\leq\textrm{VMD}_i(\mathbf{v}^*_k)\leq\delta_{i,1},
\end{align*}
which completes the proof.
\begin{remark}
Note that when at least one wheelset has no IO, \eqref{EquVMDNIC} can always be satisfied for the other wheelsets. However, when all the wheelsets have IOs simultaneously, \textit{Theorem \ref{ThmVMDNIC}} indicates that the wheelset with the slightest IO is not isolable by the VMD index. To overcome this problem, a virtual wheelset, whose velocity is calculated based on the velocity and inertia of the EMU, is always introduced in real-world usage of the VMD index. It can be seen that after employing the virtual wheelset, \eqref{EquVMDNIC} holds for all the real wheelsets.
\end{remark}

\begin{theorem}\label{ThmVMDNDC}
For the VMD index, a necessary detectability condition of the WIO is
\begin{align}\label{EquVMDNDC}
\Xi_k\mathbf{f}_k\neq f_k\mathbf{1}_p,\quad\forall f_k\in{\mathbb R}^{1}.
\end{align}
\end{theorem}
\textbf{Proof.} Note that if $\forall i\in\{1,\cdots,p\},\textrm{VMD}_i(\Xi_k\mathbf{f}_k)=0$, then the WIO is undetectable.
This means that if the WIO is detectable, then
\begin{align}\label{EquVMDNDCm1}
\exists i\in\{1,\cdots,p\},\ \textrm{such that}\ \textrm{VMD}_i(\Xi_k\mathbf{f}_k)\neq 0,
\end{align}
which is equivalent to \eqref{EquVMDNDC} because of \eqref{EquVMDProperty1}.
\begin{remark}
\textit{Theorem \ref{ThmVMDNDC}} means that a WIO which simultaneously affects all the wheelsets to the same degree will not be detected by the VMD index. Fortunately, by introducing a virtual wheelset, \eqref{EquVMDNDC} is always satisfied and the problem is readily solved.
\end{remark}

\begin{theorem}\label{ThmVMDSIC}
For the VMD index, a sufficient isolability condition of the $i$th wheelset's IO is
\begin{align}\label{EquVMDSIC}
\textrm{VMD}_i(\Xi_k\mathbf{f}_k)>\delta_{i,1}+\phi_{i,1}.
\end{align}
\end{theorem}
\textbf{Proof.} A sufficient isolability condition should guarantee that
\begin{align}\label{VMDSICRequirement}
\textrm{VMD}_i(\mathbf{v}^f_k)=\textrm{VMD}_i(\mathbf{v}^*_k+\Xi_k\mathbf{f}_k)>\delta_{i,1}.
\end{align}
Following properties of the $\textrm{VMD}_i()$ operator, we have
\begin{align}\label{VfVMDGeq}
\textrm{VMD}_i(\mathbf{v}^f_k)&=\textrm{VMD}_i\left( \Xi_k\mathbf{f}_k-(-\mathbf{v}^*_k) \right)\nonumber\\
&\geq\textrm{VMD}_i(\Xi_k\mathbf{f}_k)-\textrm{VMD}_i(-\mathbf{v}^*_k).
\end{align}
Then, by incorporating \eqref{VMDNorConstraints}, \eqref{VMDSICRequirement} and \eqref{VfVMDGeq}, we obtain \eqref{EquVMDSIC}.
\begin{remark}
To interpret the derived sufficient isolability condition \eqref{EquVMDSIC} more clearly, let us take a single wheelset IO for example. Without loss of generality, suppose the IO occurs on the $i$th wheelset. Under this circumstance, we have $\Xi_k\mathbf{f}_k=\mathbf{e}_{pi}f_k$, and then \eqref{EquVMDSIC} reduces to
\begin{align}
f_k>\delta_{i,1}+\phi_{i,1},
\end{align}
which means that if the magnitude of a single wheelset IO is larger than $\delta_{i,1}+\phi_{i,1}$, the WIO can be successfully isolated.
\end{remark}

\subsection{Analysis for the WMA-VMD index}\label{DIanalysisWMAVMDSub}
In this subsection, the above obtained results are generalized to the WMA-VMD index, under the assumption that $\mathbf{a}^{*}_W$ is positive, i.e., $\forall m\in\{1,2,\cdots,W\},\ a^*_m>0$. We make the assumption because we find it always true in this specific application of WIO detection and isolation. Practical running data of EMUs (see Section \ref{ExperimentSec}) show that the obtained covariance matrix $\hat{\mathbf{\Gamma}}^W_i$ satisfies the conditions of the OWV being positive in \textit{Theorem \ref{ThmOptimalWeightPositive}}.
Thus, hereafter, we assume that the OWV is positive.

We also use the WIO model \eqref{IOModel} here. Since $\mathbf{v}^*$ means neither slip nor slide occurs, it satisfies
\begin{align}\label{WMAVMDNorConstraints}
\widetilde{\textrm{VMD}}_{i,k}(\mathbf{a}^{*}_W,\mathbf{v}^*)\leq\delta_{i,W},\widetilde{\textrm{VMD}}_{i,k}(\mathbf{a}^{*}_W,-\mathbf{v}^*)\leq\phi_{i,W},
\end{align}
where $\delta_{i,W}$ is calculated by \eqref{PWMAVMDCL} and
\begin{align}\label{NWMAVMDCL}
\phi_{i,W}=\max_{k\in\{W,W+1,\cdots,N\}}\widetilde{\textrm{VMD}}_{i,k}(\mathbf{a}^*_W,-\mathbf{v}).
\end{align}

\begin{theorem}\label{ThmWMAVMDNIC}
For the WMA-VMD index, a necessary isolability condition of the $i$th wheelset's IO is
\begin{align}\label{EquWMAVMDNIC}
\widetilde{\textrm{VMD}}_{i,k}(\mathbf{a}^{*}_W,\Xi\mathbf{f})\neq 0,
\end{align}
where
\begin{align}\label{EquWMAVMDFault}
\widetilde{\textrm{VMD}}_{i,k}(\mathbf{a}^{*}_W,\Xi\mathbf{f})=\sum\limits_{j=1}^{W}a^*_j\textrm{VMD}_i(\Xi_{k-j+1}\mathbf{f}_{k-j+1}).
\end{align}
\end{theorem}
\textbf{Proof.} Substituting \eqref{IOModel} into the WMA-VMD index and following \eqref{VfVMDLeq}, we have
\begin{align}
\widetilde{\textrm{VMD}}_{i,k}&(\mathbf{a}^{*}_W,\mathbf{v}^f)=\sum\limits_{j=1}^{W}a^*_j\textrm{VMD}_i(\mathbf{v}^f_{k-j+1})\nonumber\\
&\leq\widetilde{\textrm{VMD}}_{i,k}(\mathbf{a}^{*}_W,\mathbf{v}^*)+\widetilde{\textrm{VMD}}_{i,k}(\mathbf{a}^{*}_W,\Xi\mathbf{f}).
\end{align}
Then, we can conclude that \eqref{EquWMAVMDNIC} is a necessary isolability condition, because otherwise, we have
\begin{align*}
\widetilde{\textrm{VMD}}_{i,k}(\mathbf{a}^{*}_W,\mathbf{v}^f)\leq\widetilde{\textrm{VMD}}_{i,k}(\mathbf{a}^{*}_W,\mathbf{v}^*)\leq\delta_{i,W}.
\end{align*}

\begin{theorem}\label{ThmWMAVMDNDC}
For the WMA-VMD index, a necessary detectability condition of the WIO is
\begin{align}\label{EquWMAVMDNDC}
\exists m\in\{1,2,\cdots,W\},&\quad\textrm{such that}\nonumber\\
\Xi_{k-m+1}\mathbf{f}_{k-m+1}\neq f_{k-m+1}&\mathbf{1}_p,\quad\forall f_{k-m+1}\in{\mathbb R}^{1}.
\end{align}
\end{theorem}
\textbf{Proof.} Since $\mathbf{a}^{*}_W$ has been assumed to be positive, according to \eqref{EquWMAVMDFault}, if the WIO is detectable, then $\exists i\in\{1,\cdots,p\}, m\in\{1,\cdots,W\}$, such that
\begin{align}
\textrm{VMD}_i(\Xi_{k-m+1}\mathbf{f}_{k-m+1})\neq 0,
\end{align}
which is equivalent to \eqref{EquWMAVMDNDC} because of \eqref{EquVMDProperty1}.

\begin{remark}
When at least one wheelset has no IO, \eqref{EquWMAVMDNIC} is always satisfied for the other wheelsets. However, when all the wheelsets have IOs simultaneously, \textit{Theorem \ref{ThmWMAVMDNIC}} indicates that the wheelset, whose IO is always slightest over a certain period of time, is not isolable by the WMA-VMD index.
\textit{Theorem \ref{ThmWMAVMDNDC}} further says that an IO affecting all the wheelsets to the same degree simultaneously over a certain period of time will not be detected by the WMA-VMD index. Fortunately, these weaknesses can also be overcome by a virtual wheelset.
\end{remark}

\begin{theorem}\label{ThmWMAVMDSIC}
For the WMA-VMD index, a sufficient isolability condition of the $i$th wheelset's IO is
\begin{align}\label{EquWMAVMDSIC}
\widetilde{\textrm{VMD}}_{i,k}(\mathbf{a}^{*}_W,\Xi\mathbf{f})>\delta_{i,W}+\phi_{i,W}.
\end{align}
\end{theorem}
\textbf{Proof.} A sufficient isolability condition should guarantee that
\begin{align}\label{WMAVMDSICRequirement}
\widetilde{\textrm{VMD}}_{i,k}(\mathbf{a}^{*}_W,\mathbf{v}^f)>\delta_{i,W}.
\end{align}
Following \eqref{VfVMDGeq}, we have
\begin{align}\label{VfWMAVMDGeq}
\widetilde{\textrm{VMD}}_{i,k}&(\mathbf{a}^{*}_W,\mathbf{v}^f)=\sum\limits_{j=1}^{W}a^*_j\textrm{VMD}_i(\mathbf{v}^f_{k-j+1})\nonumber\\
&\geq\widetilde{\textrm{VMD}}_{i,k}(\mathbf{a}^{*}_W,\Xi\mathbf{f})-\widetilde{\textrm{VMD}}_{i,k}(\mathbf{a}^{*}_W,-\mathbf{v}^*).
\end{align}
Then, by incorporating \eqref{WMAVMDNorConstraints}, \eqref{WMAVMDSICRequirement} and \eqref{VfWMAVMDGeq}, we obtain \eqref{EquWMAVMDSIC}.

\begin{remark}
We also use the above mentioned single wheelset IO case for demonstration. At this time, \eqref{EquWMAVMDSIC} reduces to
\begin{align}
\tilde{f}_k>\delta_{i,W}+\phi_{i,W},\quad\tilde{f}_k=\sum\limits_{j=1}^{W}a^*_jf_{k-j+1}.
\end{align}
This means that if the average magnitude of a single wheelset IO over a certain period of time is larger than $\delta_{i,W}+\phi_{i,W}$, the IO can be successfully isolated.
\end{remark}

\begin{theorem}\label{ThmCL1vsW}
For a fixed $i$, we have $\delta_{i,1}\geq\delta_{i,W}$ and $\phi_{i,1}\geq\phi_{i,W}$.
\end{theorem}
\textbf{Proof.} According to \eqref{PVMDCL} and \eqref{NVMDCL}, $\forall k\in\{1,\cdots,N\}$, we have
\begin{align}
\textrm{VMD}_i(\mathbf{v}_k)\leq\delta_{i,1},\quad \textrm{VMD}_i(-\mathbf{v}_k)\leq\phi_{i,1}.
\end{align}
Thus, according to \eqref{WMAVMDv}, we have
\begin{align*}
\widetilde{\textrm{VMD}}_{i,k}(\mathbf{a}^*_W,\mathbf{v})&\leq\delta_{i,1}\sum\limits_{j=1}^{W}a^*_j=\delta_{i,1},\\
\widetilde{\textrm{VMD}}_{i,k}(\mathbf{a}^*_W,-\mathbf{v})&\leq\phi_{i,1}\sum\limits_{j=1}^{W}a^*_j=\phi_{i,1}.
\end{align*}
Note that here we use the assumption that the OWV is positive. Then, following \eqref{PWMAVMDCL} and \eqref{NWMAVMDCL}, we derive $\delta_{i,1}\geq\delta_{i,W}$, $\phi_{i,1}\geq\phi_{i,W}$.

\begin{remark}\label{ConditionWMAVMDincreaseW}
Note that \eqref{EquWMAVMDNIC} and \eqref{EquWMAVMDNDC} are necessary conditions of \eqref{EquVMDNIC} and \eqref{EquVMDNDC}, respectively. This, together with \textit{Theorem \ref{ThmCL1vsW}}, reveals that \eqref{EquWMAVMDNIC}, \eqref{EquWMAVMDNDC} and \eqref{EquWMAVMDSIC} are easier to be satisfied than \eqref{EquVMDNIC}, \eqref{EquVMDNDC} and \eqref{EquVMDSIC}, respectively. It demonstrates that the WMA-VMD index has advantage over the VMD index.
Additionally, since $\hat{S}_i(\mathbf{a}^*_{W-1})=\hat{S}_i([\mathbf{a}^*_{W-1};0])\geq\hat{S}_i(\mathbf{a}^*_{W})$, we can conclude that the variance of the WMA-VMD index is a non-increasing function of $W$. Thus, for a fixed $i$, $\delta_{i,W}$ and $\phi_{i,W}$ have non-increasing trends with respect to $W$.
This, together with the fact that \eqref{EquWMAVMDNIC} and \eqref{EquWMAVMDNDC} with window length $W$ are necessary conditions of themselves with window length $W-1$ respectively, reveals that \eqref{EquWMAVMDNIC}, \eqref{EquWMAVMDNDC} and \eqref{EquWMAVMDSIC} are easier to be satisfied as $W$ gets larger.
\end{remark}

\subsection{Selection of the window length}\label{SelectWinSub}
As discussed in \textit{Remark \ref{ConditionWMAVMDincreaseW}}, larger $W$ makes the WMA-VMD index more sensitive to WIOs.
However, a $W$ larger than the duration of WIOs may reduce the sensitivity, since the left side of \eqref{EquWMAVMDSIC} will decrease when samples with no WIO are also included in the time window.
Therefore, the selection of window length should consider both the duration and magnitude of WIOs.

In practice, in order to avoid reducing traction or braking force frequently, some minor WIOs are allowed.
Denote $\check{f}_i$ as the magnitude of the $i$th wheelset's maximum tolerable IO, which can be known from engineering experience and expert knowledge.
Then, we suggest choosing the smallest window length that guarantees the isolation of WIOs larger than $\check{f}_i$, i.e.,
\begin{align}
W^*=\arg\min_{W}\ \check{f}_i>\delta_{i,W}+\phi_{i,W}.
\end{align}
Because in this case, by the use of a virtual wheelset, the sufficient isolability condition \eqref{EquWMAVMDSIC} holds:
\begin{align}
\widetilde{\textrm{VMD}}_{i,k}(\mathbf{a}^{*}_{W^*},\Xi\mathbf{f})\geq\check{f}_i\sum\limits_{j=1}^{W^*}a^*_j>\delta_{i,W^*}+\phi_{i,W^*}.
\end{align}

\section{Experiment}\label{ExperimentSec}
In this section, experimental studies are carried out on a hardware-in-the-loop (HIL) platform to demonstrate the effectiveness of the proposed WMA-VMD index. Without loss of generality, DI of WIOs on a motor car of an EMU set are studied.

\subsection{Data acquisition and hardware-in-the-loop platform}
In the experimental studies, we collect training, validation and test datasets, respectively. The training dataset is used to determine the OWV, the window length and the corresponding CL. The validation and test datasets are used to examine the WIO detection and isolation performance of the developed methods. Note that both the training and validation datasets should be WIO-free, while the test dataset should be injected with WIOs.
The training and validation samples provided by CRRC Zhuzhou Institute Co., Ltd., are the practical running data of a WIO-free EMU.
These two datasets, which totally contain 249193 samples, are nearly 24 hours' continuous records of the WIO-free EMU's wheelset velocities (driving wheelsets). Part of the validation dataset is displayed in Fig.~\ref{PartTrainData} and explained in Section \ref{VMDindexSub}.

\begin{figure}
\centering\includegraphics[width=1\linewidth]{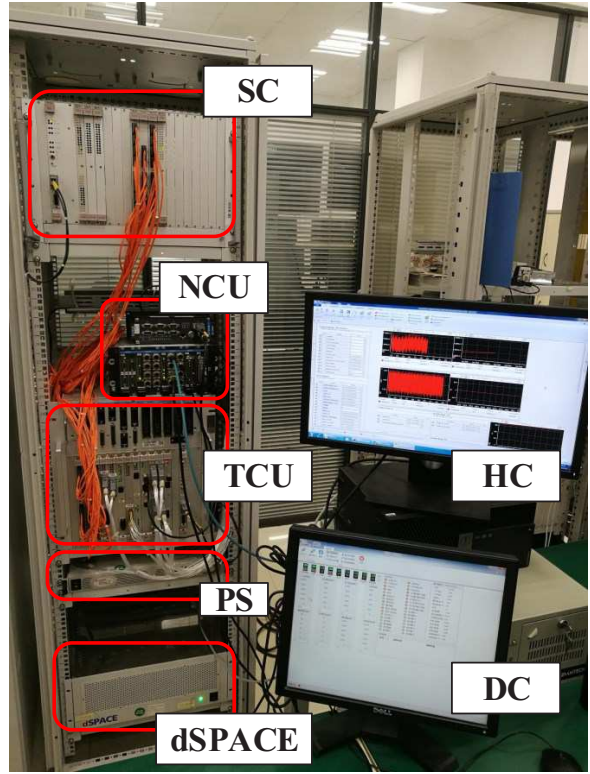}
\caption{Hardware-in-the-loop platform of EMUs.} \label{HILplatform}
\end{figure}

The test samples are collected from an HIL platform of EMUs as shown in Fig.~\ref{HILplatform}.
The HIL platform, which is jointly developed by CRRC Zhuzhou Institute Co., Ltd. and Central South University \cite{Yang2018Hardware}, consists of a simulated driver console (DC), a dSPACE real-time simulator, a power source (PS), a traction control unit (TCU), a network control unit (NCU), a signal conditioner (SC), and a host computer (HC).
Through the simulated DC, various control operations such as traction and braking can be carried out in real time during the experiments.
The wheelset and EMU dynamics, and the EMU electrical drive system (traction transformer, traction converter and traction motor), are all modeled in dSPACE. Model parameters are set to the same as nominal parameters of the CRH2-type EMU. Moreover, the TCU, NCU and SC in the HIL platform are real devices used in CRH2-type EMUs. Therefore, the HIL platform can provide a realistic simulation of the actual running of EMUs.
In addition, supporting software developed by Central South University has been installed on the HC. In this way, real-time fault injection and system monitoring, data acquisition and storage, and diagnosis algorithm evaluation can be accomplished via the HC.

In such an HIL platform of EMUs, intermittent slips can be injected by changing the adhesion coefficient intermittently (e.g., switching the WR surface condition of some driving wheelsets between dry and wet, repeatedly) during the traction period \cite{Wang2016Locomotive,Diao2017Taking}. Intermittent slides can be injected in the same way during the braking period. Experiments based on the HIL platform are conducted to inject WIOs and obtain the test dataset via the HC. The injected WIOs are on the wheelsets of a motor car of an EMU set. Motor cars are propelled by four independently driven wheelsets. The sampling interval for the test dataset is consistent with that for training and validation datasets.

\subsection{Experimental results and discussions}
After data pre-processing, the training samples are used to calculate the OWV through Algorithm 1.
Calculation results show that the obtained OWVs are all positive. Therefore, the detectability and isolability analyses given in Sections \ref{DIanalysisVMDSub} and \ref{DIanalysisWMAVMDSub} are applicable here.
According to Section \ref{SelectWinSub}, by considering the duration and magnitude of tolerable WIOs, the window length is selected as $W=3$.
\begin{figure} %Traction or coasting mode (20-324, 600-823), braking mode (325-500, 824-983)
\centering\includegraphics[width=1\linewidth]{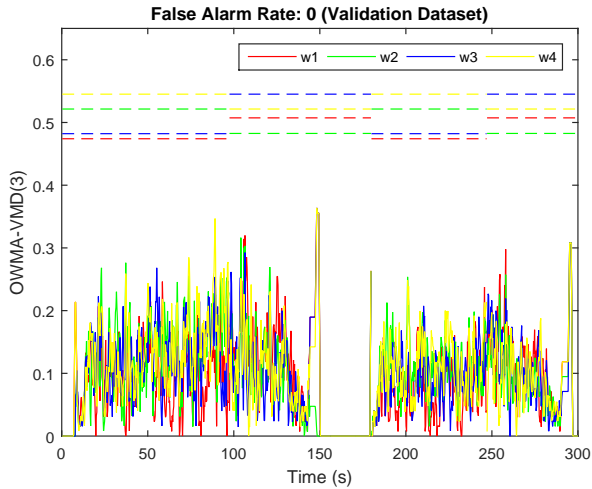}
\caption{Detection and isolation results on the validation dataset.} \label{DIvalueResults}
\end{figure}

Since false alarms cause unnecessary reduction in traction and braking force, we use the validation dataset to examine the false alarms of the proposed WMA-VMD index. Note that in traction or coasting mode (i.e., 6-97.2s and 180-246.9s), $\widetilde{\textrm{VMD}}_{i,k}(\mathbf{a}^*_W,\mathbf{v}^f)$ is used, while in braking mode (i.e., 97.2-150s and 246.9-294.9s), $\widetilde{\textrm{VMD}}_{i,k}(\mathbf{a}^*_W,-\mathbf{v}^f)$ is used. DI results on the validation dataset are shown in Fig.~\ref{DIvalueResults}, where solid lines and dashed lines represent four wheelsets' WMA-VMD indices and their corresponding CLs, respectively. It can be seen that there is no false alarm. This is because for each wheelset, the CL is chosen as the largest WMA-VMD index of all the training samples. This way of determining the CL can effectively reduce unnecessary false alarms \cite{Zhou2018Fault}.
Note that in the same running mode, the CLs can be different, because even though training samples are the same, the OWVs, as well as the $\textrm{VMD}_i(\mathbf{v})$ values are different for the four wheelsets.
Moreover, for the same wheelset, CLs in traction mode and braking mode can also be different. This is because the OWVs in these two modes, as well as the values of $\textrm{VMD}_i(\mathbf{v})$ and $\textrm{VMD}_i(-\mathbf{v})$, are different.
\begin{figure}
\centering\includegraphics[width=1\linewidth]{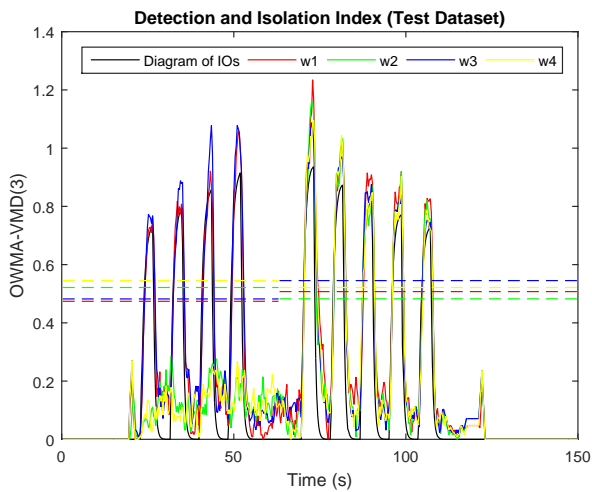}
\caption{Detection and isolation indices on the test dataset.} \label{DItestIndex}
\end{figure}

\begin{figure}
\centering\includegraphics[width=1\linewidth]{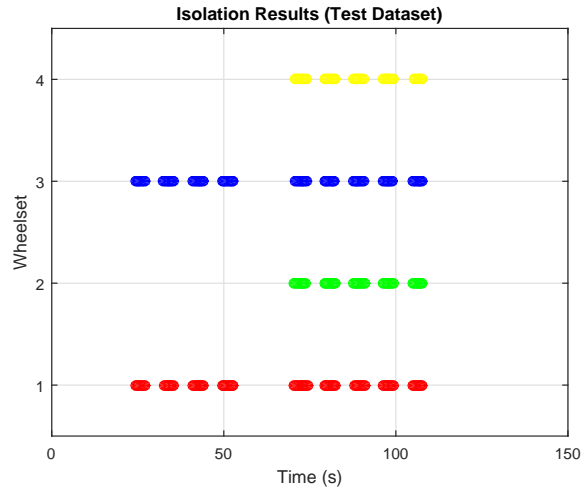}
\caption{Isolation results on the test dataset.} \label{ItestResults}
\end{figure}

In the test dataset, the EMU is in traction mode at first. Then, from 63s, the EMU starts to brake until it stops.
Intermittent slips are injected on both the first and third wheelsets during the traction period, while intermittent slides are injected on all of the four wheelsets during the braking period. Note that since the TCU used in the experiments is a real device of the CRH2-type EMU, it will alarm if the injected WIOs can be detected by current CRH2-type EMU's ASS strategies. A diagram of the injected WIOs, the WMA-VMD indices and their corresponding CLs are shown in Fig.~\ref{DItestIndex} with a black line, solid lines and dashed lines, respectively. In addition, DI results are given in Fig.~\ref{ItestResults}. It is observed that the proposed WMA-VMD index can detect and isolate the appearing and disappearing of WIOs effectively, whereas current ASS strategies as given in Section \ref{CurrentStrategySub} do not raise any alarm.

Sufficient isolability conditions of wheelset intermittent slips or slides using different window lengths, i.e., values of $\delta_{i,W}+\phi_{i,W}$ in \eqref{EquWMAVMDSIC}, are given in Table \ref{TableSIC} (rounded to four decimals). It can be seen that when the used window length is same, the sufficient isolability conditions for the four wheelsets, or for slip and slide are similar. This is because wheelsets on the same EMU car are often of the same type, size and wearing degree. Moreover, it is observed that the minimum magnitude of isolable WIOs has a decreasing trend as the window length increases. By employing the WMA technique ($W\!=\!3$), DI of a WIO as small as 1km/h, which is nearly 66\% magnitude of the isolable WIO without WMA ($W\!=\!1$), can be achieved. This demonstrates the advantage of the WMA technique.
\begin{table}\centering
\caption{Sufficient isolability conditions using different $W$}
\begin{tabular}{c|c|c|c|c}\hline
Slip (km/h) & $i=1$ & $i=2$ & $i=3$ & $i=4$ \\\hline
$W=1$ & 1.5644 & 1.6118 & 1.5644 & 1.7303\\\hline
$W=2$ & 1.1259 & 1.1022 & 1.1140 & 1.1496\\\hline
$W=3$ & 0.9822 & 1.0053 & 1.0273 & 1.0666\\\hline
Slide (km/h) & $i=1$ & $i=2$ & $i=3$ & $i=4$ \\\hline
$W=1$ & 1.5644 & 1.6118 & 1.5644 & 1.7303\\\hline
$W=2$ & 1.1259 & 1.1022 & 1.1140 & 1.1496\\\hline
$W=3$ & 0.9814 & 1.0040 & 1.0275 & 1.0666\\\hline
\end{tabular}
\label{TableSIC}
\end{table}

\section{Conclusions}\label{ConclusionSec}
In this paper, the detection and isolation (DI) performance of wheelset intermittent over-creeps (WIOs) has been improved using only velocity
measurements. A variable-to-minimum difference (VMD) index has been combined with a weighted moving average (WMA) technique to form a WMA-VMD index. The WMA-VMD index is well suited to cases where measurement variables are similar to each other under normal conditions, e.g., the DI task of WIOs. Using the VMD index, the DI task is accomplished in one step, and a non-stationary process is transformed into a stationary process, which is easier to be monitored. Using the WMA technique, the robustness and sensitivity of the VMD index are enhanced by employing a time window and a weight vector. Compared with the MA technique, WMA can use the correlation information to further increase the index's DI capability by finding an optimal weight vector (OWV).

The uniqueness of the OWV for the WMA-VMD index has been proven, and properties of the OWV have been revealed. We have found that the OWV possesses a symmetrical structure, and the equally weighted scheme is optimal when process data exhibit no autocorrelation. These verify the optimality of existing MA-based DI methods when applied to independent data. Moreover, by analyzing the properties of two nonlinear, discontinuous operators, $\min$ and $\textrm{VMD}_i$, the necessary conditions and the sufficient conditions for WMA-VMD-based detectability and isolability of WIOs have been derived. The effectiveness of the developed methods has been demonstrated by experimental studies using practical running data and a hardware-in-the-loop platform of an EMU.

%The developed methods have been evaluated by experimental studies using practical running data and a hardware-in-the-loop platform of an EMU. Experimental results show that the developed methods are effective.
%Experimental results show that the developed methods can detect and isolate WIOs as small as 1km/h, which is 30\% of the original detectable WIOs.
%One of the drawback is highly depend on the accuracy of sensors. not applicable to low speed range.

\bibliographystyle{IEEEtranTIE}
\bibliography{arXiv_OcDI}
\end{document}